\documentclass[10pt,a4paper]{article}
\pdfoutput=1
\usepackage{jheppub}
\usepackage{arydshln}
\usepackage{tikz}
\usetikzlibrary{positioning}
\usepackage{lipsum}
\usepackage{parskip}
\usepackage{xcolor}
\usepackage[all]{xy}
\usepackage{amsfonts}
\usepackage{amscd}
\usepackage{amssymb}
\usepackage{amsmath,bbm}
\usepackage{graphicx}
\usepackage{epsfig}
\usepackage{latexsym}
\usepackage{mathtools}
\usepackage{hyperref}
\usepackage{amsthm}
\usepackage[vcentermath]{youngtab}
\usepackage{accents}
    
\usepackage{calligra}

\DeclareMathAlphabet{\mathcalligra}{T1}{calligra}{m}{n}
\DeclareFontShape{T1}{calligra}{m}{n}{<->s*[2.2]callig15}{}

\def \be  {\begin{equation}}
\def \ee  {\end{equation}}
\def \bea {\begin{equation}\begin{aligned}}
\def \eea {\end{aligned}\end{equation}}
\def \ba  {\begin{eqnarray}}
\def \ea  {\end{eqnarray}}
\def \bb  {\begin{thebibliography}}
\def \eb  {\end{thebibliography}}
\def \lab #1 {\label{#1}}

\newcommand\tr{\mathrm{Tr}}

\definecolor{cardinal}{rgb}{0.6,0,0}
\definecolor{darkgreen}{rgb}{0,0.5,0}
\definecolor{golden}{rgb}{0.92, 0.7, 0}
\definecolor{midnight}{rgb}{0, 0, 0.5}
\definecolor{darkblue}{rgb}{0.2, 0, 0.8}

\theoremstyle{definition}

\def\CN{{\cal N}}
\def\CS{{\cal S}}
\def\CV{{\cal V}}

\newif\ifniklas\niklastrue
\RequirePackage{verbatim}

\usepackage{booktabs}
\usepackage{multirow}
\usetikzlibrary{cd}

\usetikzlibrary{arrows}
\usetikzlibrary{positioning}
\usetikzlibrary{decorations.pathmorphing}
\usetikzlibrary{decorations.markings}
\def\arrowhead{angle 90}
\tikzset{>=\arrowhead}
\pgfarrowsdeclaredouble{<<}{>>}{\arrowhead}{\arrowhead}
\pgfarrowsdeclaretriple{<<<}{>>>}{\arrowhead}{\arrowhead}
\pgfarrowsdeclaredouble{<<<<}{>>>>}{<<}{>>}
\tikzstyle{W}=[draw, circle, minimum width=0.6cm, scale=1]
\def\dist{2}
\tikzstyle{B}=[draw,circle,fill=black,scale=1]
\tikzstyle{D}= [circle, minimum size=1em]
\tikzstyle{H}=[draw,circle,fill=gray,scale=1]
\tikzstyle{R}=[draw, inner sep=4pt, minimum size=1.8em]
\tikzstyle{every picture}=[scale=1,baseline=(current bounding box.south)]

\makeatletter

\def\CN{\mathcal{N}}
\def\CS{\mathcal{S}}
\def\CV{\mathcal{V}}

\def\II{I\!\!I\!}

\renewcommand{\tilde}[1]{\widetilde{#1}}
\renewcommand{\hat}[1]{\widehat{#1}}

\usepackage{subfiles}

\title{Macdonald index from 3d TQFT}
\author[a]{Heeyeon Kim,}
\author[a]{Hongseok Kim,}
\author[a, b]{and Jaewon Song}
\affiliation[a]{Department of Physics, Korea Advanced Institute of Science and Technology \\291 Daehak-ro, Yuseong-gu, Daejeon 34141, Republic of Korea}
\affiliation[b]{Walter Burke Institute for Theoretical Physics, California Institute of Technology\\
Pasadena, CA 91125, USA}
\emailAdd{heeyeon.kim@kaist.ac.kr}
\emailAdd{rstone@kaist.ac.kr}
\emailAdd{jaewon.song@kaist.ac.kr}

\abstract{We propose a new fermionic sum formula for the Macdonald index of a class of Argyres-Douglas theories.
The formula arises naturally from a three-dimensional topological field theory obtained via a twisted dimensional reduction of the 4d theory. Such a reduction often gives rise to a 3d $\CN=2$ abelian Chern-Simons matter theory, which is expected to flow to an $\CN=4$ superconformal fixed point. After performing a topological twist, we obtain a 3d TFT admitting boundary conditions that support the vertex operator algebra associated with the original 4d theory. 
In this framework, the Macdonald index appears as a half-index of the 3d gauge theory, with the Macdonald grading determined by a distinguished $U(1)_A$ symmetry in the infrared $\CN=4$ superconformal algebra. We present a general procedure to identify this $U(1)_A$ symmetry and, whenever possible, show that it reproduces the refined character of the associated vertex operator algebra, thereby recovering the Macdonald index.   
Our construction also gives a hint towards the IR formula for the Macdonald index in terms of 4d BPS particles on the Coulomb branch.
}

\begin{document}
	
\maketitle

\section{Introduction}

The correspondence between four-dimensional $\CN=2$ superconformal field theories (SCFTs) and two-dimensional vertex operator algebras (VOAs), first proposed in \cite{Beem:2013sza}, establishes a precise relation between protected operator sectors of the 4d theory and two-dimensional chiral algebras. In this framework, the Schur index of a 4d $\CN=2$ SCFT, defined as
\begin{align}
I_S(q) = \text{tr }(-1)^F q^{\Delta-R}\ ,
\end{align}
plays a central role, as it is identified with the vacuum character of the associated VOA. Beyond this algebraic relation, the Schur index also admits a rather mysterious infrared reformulation, proposed in \cite{Cordova:2015nma,Cordova:2016uwk}, in which it is expressed as
\begin{align}\label{quantum monodromy intro}
I_S(q) =(q)_{\infty }^{2r} \text{Tr }{\mathcal O}(q)\ ,
\end{align}
where $r$ is the rank of the Coulomb branch and ${\mathcal O}(q)$ is the quantum monodromy operator constructed from the BPS 
spectrum in the Coulomb branch of the 4d theory. This representation provides a deep connection between UV operator algebras and infrared BPS dynamics, suggesting interrelations among wall-crossing phenomena, protected operator algebras, and the underlying modular structure of supersymmetric field theories.

This intriguing relation between the chiral algebra and the infrared BPS spectrum can be better understood from a three-dimensional viewpoint. In particular, as illustrated in \cite{Dedushenko:2023cvd}, for Argyres-Douglas type theories whose operator spectrum includes fractional $U(1)_r$-charges, one can consider an $r$-twisted compactification that gives rise to a 3d $\CN=4$ SCFT \cite{Gang:2018huc, Gang:2023rei}. After performing a topological twist, we obtain a three-dimensional topological field theory (TFT) whose boundary supports the 2d VOA associated with the original 4d SCFT. 

Although the resulting 3d SCFTs are generally non-Lagrangian, there exists a natural way to obtain an effective Lagrangian description with $\CN=2$ symmetry from the Coulomb branch perspective of the 4d theory. As discussed in \cite{Gaiotto:2024ioj} (see also \cite{Kim:2024dxu, Go:2025ixu}), by focusing on the BPS spectrum that remains stable under the $r$-twisted circle compactification, we are naturally led to a 3d $\CN=2$ abelian Chern-Simons matter theory subject to a superpotential deformation, which is expected to flow to the non-Lagrangian SCFT with $\CN=4$ supersymmetry enhancement. This description allows us to compute various observables for the 3d SCFT, including the half-index of the A-twisted theory, defined by
\begin{align}
\II^A(q) = \mathrm{tr}_B~(-1)^F q^{J+R_H}\ ,
\end{align}
where the trace is taken over the space of boundary local operators, and $R_{H},~ R_C$ correspond to Cartans of the $SU(2)_H \times SU(2)_C$ R-symmetry of the $\CN=4$ algebra.
For a large class of Argyres-Douglas (AD) theories, the $A$-twisted half-index obtained in this way is identified with the trace of the quantum monodromy operator \eqref{quantum monodromy intro}, which in turn provides a physical motivation for the infrared trace formula of the Schur index. See also \cite{ArabiArdehali:2024ysy,Ardehali:2024} for an alternative approach based on studying the high-temperature limit of the superconformal index. 

In this work, we revisit this story by introducing a one-parameter refinement of the Schur index, namely the Macdonald index, defined by
\begin{align}\label{Macdonald intro}
I_\text{Mac}(q,T) = \mathrm{tr}_M(-1)^F q^{\Delta-R}T^{R-r}\ ,
\end{align}
where the trace is over the operators satisfying the relation $\Delta = 2R+2j_2-r$ and $j_1-j_2+r=0$. This is a natural refinement to consider in the context of the 4d SCFT/2d VOA correspondence, since it is built from the same protected operator spectrum as the Schur index while keeping track of an additional $U(1)$ charge. From the perspective of the associated VOA, this refinement arises from the existence of a natural filtration structure, as first discussed in \cite{Song:2016yfd}. See also \cite{Agarwal:2018zqi, Bonetti:2018fqz, Beem:2019tfp, Xie:2019zlb, Agarwal:2021oyl} for related development.

The 3d $\CN=4$ SCFT obtained from the twisted circle reduction has an $\CN=2$ superconformal subalgebra where the $U(1)_R$ symmetry embeds into $SU(2)_H\times SU(2)_C$, leaving the combination $U(1)_A = U(1)_H- U(1)_C$ as a distinguished global symmetry commuting with the $U(1)_R$. This leads to the refined twisted half-index,
\begin{align}\label{refined half intro}
\II^A(q,T) = \mathrm{tr}_B~(-1)^F q^{J+R_H} T^{R_H-R_C}\ ,
\end{align}
which can be naturally identified with the Macdonald index under the twisted compactification.

This paper aims to propose a new formula for the Macdonald index of $(A_1,G)$ type Argyres-Douglas theories \cite{ Argyres:1995jj, Argyres:1995xn, Eguchi:1996vu, Cecotti:2010fi, Xie:2012hs} from the perspective of 3d SCFTs. Our starting point is the 3d $\CN=2$ abelian Chern-Simons matter (ACSM) theory description suggested by the infrared trace formula \eqref{quantum monodromy intro}. Given such a Lagrangian description, together with the known Coulomb/Higgs branch symmetries of 3d $\CN=4$ SCFTs inherited from the parent 4d SCFT, one can often identify a natural candidate for the $U(1)$ global symmetry in the ACSM description that flows to the $U(1)_A$ symmetry in the infrared. This identification is guided by the requirement that the $\CN=2$ ACSM description admits a superpotential deformation consistent with the Coulomb/Higgs branch symmetries of the moduli space. The proposed identification of the $U(1)_A$ symmetry can be further supported, whenever possible, by $F$-maximization together with the computations of the Higgs/Coulomb branch Hilbert series.

The refined half-index \eqref{refined half intro} can then be systematically computed using the ACSM description with the fugacity for the $U(1)_A$ turned on, and it can be naturally expressed as a refined fermionic sum formula. When compared with the Macdonald index derived from 4d constructions, this formulation suggests new conjectural $q$-series identities, encoding a nontrivial correspondence between the two perspectives.

\emph{Note Added}: While finalizing this paper, we became aware of \cite{Andrews:2025tko}, which has an overlap with the result presented in section \ref{sec: conjectural Macdonald}.

\section{Macdonald index formula}

In this section, we introduce the Macdonald index, review several methods of computation, and present our conjectural closed-form formula for a number of Argyres-Douglas theories. 

\subsection{Definition of the Macdonald index}
The superconformal index counts certain BPS states with weight of $(-1)^F$ or, equivalently, short multiplets modulo recombinations \cite{Kinney:2005ej}. Because of the Bose-Fermi cancellation, this quantity is invariant under marginal deformations.  
The trace formula for the full superconformal index of a 4d $\mathcal{N}=2$ SCFT is defined as 
\begin{align}
    I(p, q, t) = \text{tr } (-1)^F p^{j_2 - j_1 - r} q^{j_2 + j_1 - r} t^{R+r} \ , 
\end{align}
where the trace is taken over the states satisfying $\Delta = 2R + 2j_2 - r$, counting $\frac{1}{8}$-BPS states. Here $(j_1, j_2)$ are the Cartans for the Lorentz group $SU(2)_1 \times SU(2)_2$. $R$ and $r$ denote the Cartan of the $SU(2)_R$ symmetry and $U(1)_r$ symmetry respectively. 
There are various simplifying limits of the 4d $\CN=2$ superconformal index. We are in particular interested in the Macdonald index, which is defined by taking $p=0$ from the above formula \cite{Gadde:2011uv}. Then the trace formula becomes
\begin{align}
    I_\text{Mac} (q, t) = \mathrm{tr}_M (-1)^F q^{\Delta - 2R + r} t^{R-r} = \mathrm{tr}_M (-1)^F q^{\Delta - R} T^{R-r} \ , 
\end{align}
    where $t=qT$ and the trace is taken over states satisfying $\Delta = 2R + 2j_2 - r$ and $j_1-j_2+r = 0$, counting $\frac{1}{4}$-BPS states. 

One can further simplify the index by taking $t=q$ or $T=1$ to get
\begin{align}
    I_\text{Schur} (q) = \text{tr } (-1)^F q^{\Delta - R} \ . 
\end{align}
This is referred to as the Schur index \cite{Gadde:2011ik}, which plays a pivotal role in many developments in the field. In particular, the Schur index computes the (super-)character of the associated VOA, in the context of 4d SCFT/VOA correspondence \cite{Beem:2013sza}. 
One can also directly obtain the Schur index by taking the $q=t$ limit from the full index. In this limit, the $p$-dependence drops out \cite{Gadde:2011uv}. 
One interesting aspect of the Schur index is that, unlike the Macdonald index, the trace can be taken over \emph{all} states in the radially quantized Hilbert space, without restricting to a certain BPS subspace.

\subsection{Computation of Macdonald index}
For a Lagrangian theory, the superconformal index can be computed simply by combining contributions from the free fields and projecting them to the gauge-invariant sector by integrating over the gauge group. 
However, for a non-Lagrangian theory, such as the Argyres-Douglas theory \cite{Argyres:1995jj, Argyres:1995xn, Eguchi:1996vu, Cecotti:2010fi, Xie:2012hs}, we need to resort to alternative methods. 
The Macdonald index for Argyres-Douglas theories can be computed using several methods, each with its own strengths and weaknesses. 
Moreover, each method yields a different formula, though the final expressions are expected to coincide, leading to non-trivial mathematical identities.

\paragraph{$\CN=1$ Lagrangian flowing to $\CN=2$ AD theory}
The AD theories of $(A_1, A_n)$, $(A_1, D_n)$, $(A_{k-1}, A_{kn-1})$, $(A_{k-1}, D_{kn+1})$ type and a few more admit $\CN=1$ Lagrangian gauge theory descriptions that flow to $\CN=2$ fixed points \cite{Maruyoshi:2016tqk, Maruyoshi:2016aim, Agarwal:2016pjo, Agarwal:2017roi, Benvenuti:2017bpg}. Therefore, one can use the contour integral formula for the full index and take the Macdonald limit to obtain the index. So far, this is the only known way to obtain the full index of AD theories. 

However, this method has some drawbacks: 
Firstly, not all AD theories admit such a Lagrangian description. For example, $(A_{k-1}, A_{n-1})$ theory with $(k, n)=1$, whose associated VOA is given by the $W$-minimal model $W(k, k+n)$, does not have a known Lagrangian description for $k>2$. 
Secondly, this method is computationally inefficient for obtaining the Macdonald index.
In particular, the integrand is singular as we take the $p\to 0$ limit.
So one must first evaluate the integral of the full index and then take the Macdonald limit. 

\paragraph{Class $\mathcal{S}$ description and TQFT structure}
Class $\CS$ theory refers to 4d $\CN=2$ SCFTs obtained via compactifying 6d $\CN=(2, 0)$ theory on a Riemann surface \cite{Witten:1997sc, Gaiotto:2009we, Gaiotto:2009hg}. Such a 4d theory is determined by the choice of 6d theory ($\Gamma \in ADE$) and a UV curve $\mathcal{C}_{g, n}$ of genus $g$ and $n$ punctures. In addition, one should prescribe a local structure for each puncture. Many of the AD theories can be realized in class $\CS$ \cite{Xie:2012hs, Wang:2015mra}. 

The superconformal index of class $\CS$ theories can be thought of as a partition function or correlation function of 2d TQFT on a UV curve \cite{Gadde:2009kb, Gadde:2011ik, Gadde:2011uv}. Such a TQFT structure originates from the S-duality and the deformation invariance of the index. 
For the AD theories admitting a class $\CS$ realization, we can utilize this correspondence. Such theories are realized by a sphere with an irregular puncture. Hence, we need to know the wave function corresponding to the irregular punctures. This is achieved for several interesting cases \cite{Buican:2015ina, Buican:2015tda, Song:2015wta, Buican:2017uka, Nishinaka:2018zwq, Watanabe:2019ssf}. 

For example, the TQFT description gives the Macdonald index for the $(A_1, A_{2n})$ theory as
\begin{align} \label{eq:A1A2nMac}
    I_{(A_1, A_{2n})}(q, t) = \frac{1}{(t^2; q)_\infty}\sum_{\lambda \ge 0} (-1)^{\lambda}q^{\lambda (\lambda+1)(n+\frac{3}{2})} \left(\frac{t}{q}\right)^{\lambda (n+2)} \frac{(q^{\lambda+1})_{\lambda} (t^2 q^{2\lambda})_\infty}{(t q^{\lambda})_{\lambda} (t q^{2\lambda+1})_\infty}[2\lambda]_{q, t} \ , 
\end{align}
where 
\begin{align}
    [n]_{q, t} = \sum_{i=0}^n \frac{(t; q)_i (t; q)_{n-i}}{(q; q)_i (q; q)_{n-i}} t^{\frac{n}{2}-i} \ . 
\end{align}
The $q$-Pochhammer symbol is defined as
\begin{align}
    (z)_n \equiv (z; q)_n \equiv \prod_{i=1}^n (1- z q^i) \ . 
\end{align}
For the $(A_1, D_{2n+1})$ theory, we have
\begin{align}\label{eq:A1D2n1Mac}
    I_{(A_1, D_{2n+1})} = \frac{1}{(tz^{\pm 2,0})_\infty} \sum_{\lambda \ge 0} (-1)^{\lambda} q^{\lambda(\lambda+1)(n+\frac{1}{2})} \left( \frac{t}{q} \right)^{\lambda(n+1)} \frac{(q^{\lambda+1})_\lambda (t^2 q^{2\lambda})_\infty}{(t q^{\lambda})_\lambda (t q^{2\lambda+1})_\infty} P_{2\lambda}(z) \ , 
\end{align}
where
\begin{align}
    P_{\lambda} (z) = \sum_{i=0}^{\lambda} \frac{(t; q)_i (t; q)_{\lambda-i}}{(q; q)_i (q; q)_{\lambda-i}} z^{2i - \lambda} \ , 
\end{align}
is the Macdonald polynomial for $SU(2)$. 

\paragraph{Refined character for the VOA}
One can employ the SCFT/VOA correspondence to compute the Schur index from the vacuum character of the associated vertex operator algebra \cite{Beem:2013sza}. Argyres-Douglas theories have relatively simple associated VOAs such as Virasoro/W-minimal models or affine Lie algebra \cite{Buican:2015ina, Cordova:2015nma, Xie:2016evu, Song:2017oew}. 

Since a VOA captures the entire Schur sector, it was expected that the VOA would somehow capture the Macdonald index, which counts the same set of operators. 
Indeed, it was found that the `refined character' of the vacuum module can capture the refined index \cite{Song:2016yfd}. See also \cite{Agarwal:2018zqi, Bonetti:2018fqz, Beem:2019tfp, Xie:2019zlb, Agarwal:2021oyl}.  
For the case of Virasoro algebra, we consider the following filtration of the vacuum module
\begin{align}
    \CV_0 \subset \CV_1 \subset \CV_2 \subset \CV_3 \cdots \ , 
\end{align}
where 
\begin{align}
    \CV_n = \textrm{span}\{ L_{-i_1} L_{-i_2}\cdots L_{-i_m} |0\rangle : m \le n \}/\{\textrm{null states} \} \ . 
\end{align}
From this, one can define a graded vector space 
\begin{align}
    V_{\textrm{gr}} = \bigoplus_{i=0}^\infty V_i = \CV_0 \oplus \bigoplus_{i=1}^{\infty} (\CV_i/\CV_{i-1}) \ , 
\end{align}
where $V_i \equiv \CV_i / \CV_{i-1}$. 
This allows us to define the refined character as
\begin{align}
    \chi_{\CV}^{\textrm{ref}}(q; T) = \sum_{i=0}^\infty (\tr_{V_i} q^{L_0}) T^i \ , 
\end{align}
where we dropped $q^{-\frac{c}{24}}$ for convenience. Upon identification $t=qT$, the refined character defined in this way is conjectured to give the Macdonald index of the corresponding 4d $\CN=2$ SCFT. 

A closed-form fermionic expression for the refined character of the Virasoro minimal model $M(2, 2k+3)$ (or Macdonald index of $(A_1, A_{2k})$ theory) is found in \cite{Foda:2019guo} to be
\begin{align} \label{eq:A1A2nFermion}
    I_\text{Mac}^{(A_1, A_{2k})} (q, T) = \sum_{n_1 \ge n_2 \ge \cdots n_k \ge 0} \frac{q^{n_1^2 + \cdots n_k^2 + n_1 + \cdots n_k}}{(q)_{n_1 - n_2} \cdots (q)_{n_{k-1}- n_k} (q)_{n_k}} T^{n_1 + \cdots + n_k} \ . 
\end{align}
One can check that this formula agrees with \eqref{eq:A1A2nMac} to high orders in $q$. The conjecture is that they are are identical, confirming that the refined character defined as above indeed computes the Macdonald index.

A fermionic formula for the $W(3, 7)$ minimal model is also proposed in \cite{Foda:2019guo}, which computes the Macdonald index of $(A_2, A_3) = (A_1, E_6)$ theory:
\begin{align}\label{eq:A2A3Mac}
    I_M^{(A_2, A_3)} &= \sum_{n_{1,2, 3, 4} \ge 0} \frac{q^{(n_1+n_2+n_3)^2 + (n_2+n_3)^2 + n_3^2 + n_4^2 + (n_1 + 2n_2 + 3n_3)n_4 + (n_1 + 2n_2 + 3n_3 +2n_4)}}{(q)_{n_1} (q)_{n_2}(q)_{n_3}(q)_{n_4}} T^{n_1 + 2n_2 +3n_3 + 2n_4} 
\end{align}

\paragraph{Hilbert series of the Arc space of a scheme} 
More recently, it was proposed that the Macdonald index can be computed from the arc space $J_\infty(X)$ of a certain scheme $X$ \cite{Bhargava:2023hsc, Andrews:2025krn, Kang:2025zub}, which encodes null state relations in the Schur sector. Sometimes $X$ is identical to Zhu's $C_2$-algebra obtained by reducing the associated VOA, dropping higher-derivative modes. 
The arc space can be thought of as a collection of arcs (curves) passing through each point in $X$. One can compute the bi-graded Hilbert series for $J_\infty(X)$, which produces the Schur index upon appropriate identification of fugacities. To obtain the Macdonald index, one should introduce a $T$-filtration similar to the refined character. From this, one can obtain a triply-graded coordinate ring, which allows us to recover the Macdonald index. We refer to \cite{Kang:2025zub} for more details. 

This method involves computing a Gr\"obner basis for the coordinate ring of the arc space (or jet scheme), which gives us a computational bottleneck. However, this method is sometimes more suitable for obtaining fermionic expressions for the index, as is demonstrated in \cite{bai2020quadratic, bruschek2011arc, Andrews:2025krn, Kang:2025zub}. 

\subsection{Conjectural formula for the Macdonald index from IR formula} \label{sec: conjectural Macdonald}

    It is conjectured in \cite{Cordova:2015nma} that the Schur index can be computed by evaluating the trace of an operator $\mathcal{O}$ in quantum torus algebra, known as the Kontsevich-Soibelman (KS) operator, constructed out of the BPS particle spectrum in the Coulomb branch of the 4d theory.
The spectrum of BPS particles depends on the point on the Coulomb branch and can jump across codimension-one walls of marginal stability. 
A key property of the operator $\mathcal{O}$ is that it remains invariant under wall-crossing, and therefore it defines an invariant of the 4d theory.

\begin{figure}
  \centering
  \begin{tikzpicture}
			\node[W] (1) at (0,0){1};
		 	\node[W] (2) at (2,0){2};
            \node[W] (3) at (4,0){3};
            \node[W] (4) at (6,0){4};
			\draw[->] (1)--(2);
            \draw[<-] (2)--(3);
            \draw[->] (3)--(4);
  \end{tikzpicture}
  \caption{\label{fig:A1A4quiver}
  BPS quiver for the $(A_1, A_4)$ theory in the canonical chamber}
\end{figure}
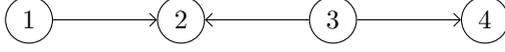

Let $\Gamma$ be the electro-magnetic charge lattice of the 4d Coulomb branch. For each charge vector $\gamma \in \Gamma$, we introduce a quantum torus algebra variable $X_{\gamma}$ which obeys
\begin{align}
	X_{\gamma}X_{\gamma'}=q^{\frac{1}{2}\left< \gamma,\gamma' \right>}X_{\gamma+\gamma'}=q^{\left< \gamma,\gamma' \right>}X_{\gamma'}X_{\gamma}.
\end{align} 
We define a KS operator as
\begin{align}
	\mathcal{O}(q) =\prod _{\gamma\in \Gamma}^{\curvearrowright}E_{q}(X_{\gamma}) \ , 
\end{align} 
where the product is taken for all BPS states of charge $\gamma\in\Gamma$, ordered so that larger $\arg(Z_{\gamma})$ appear to the right and \footnote{Notice that our definition of $E_q(z)$ is related to that of \cite{Cordova:2015nma} by replacing $q^{1/2}\rightarrow -q^{1/2}$. This choice is slightly more natural from the 3d half-index point of view with the $(-1)^F$ grading. It also leads to a more convenient assignment of $\text{Tr}[X_{\text{flavor}}]$; see e.g., eq. \eqref{flavor A odd}.}
\begin{align}
E_{q}(X)=(q^{\frac{1}{2}}X;q)_{\infty }^{-1}=\sum_{n\in\mathbb{Z}_{\geq0}}\frac{(q^{\frac{1}{2}}X)^{n}}{(q;q)_{n}}
\end{align}
is the quantum dilogarithm.
Then, the Schur index of the theory is calculated by
\begin{align}\label{eq:traceSchur}
	I_{\text{Schur}}(q)=(q)_{\infty }^{2r}\, \mathrm{Tr}\left[ \mathcal{O}(q)\right] \ , 
\end{align} 
where $r$ is the rank of the Coulomb branch.
The trace is defined by
\begin{align}
	\mathrm{Tr} [X_{\gamma}]=
	\begin{cases}
		\prod _{i}\mathrm{Tr} [X_{\gamma_{f_{i}}}]^{f_{i}(\gamma)} & \left<\gamma,\gamma'\right>=0\quad \forall \gamma'\in \Gamma\\
		0 &\text{else}
	\end{cases}.
\end{align} 
As shown in \cite{Cordova:2015nma}, this IR formula indeed reproduces the Schur index. 

For the Argyres-Douglas theories of type $(A_1,G)$, there exists the so-called canonical chamber, where all the nodes are either sink or source. In this chamber the formula reads
\begin{align}
    \mathcal{O}(q) &= \prod_{i \textrm{~odd}} E_q(X_{\gamma_i}) \prod_{j \textrm{~even}} E_q(X_{\gamma_j}) \prod_{i \textrm{~odd}} E_q(X_{-\gamma_i}) \prod_{j \textrm{~even}} E_q(X_{-\gamma_j}) \ . 
\end{align}
For example, for $(A_1,A_{2n})$ theory, we obtain a fermionic sum type expression:
\begin{align} \label{eq:A1A2nIR}
I_{\text{Schur}}(q)= (q)_\infty^{2n}\sum_{m_1, \cdots, m_{2n} \ge 0} \frac{q^{ \sum_{i=1}^{2n} m_i + \sum_{i=1}^{2n-1} m_i m_{i+1}}}{\prod_{i=1}^{2n}(q)^2_{m_i}} \ . 
\end{align}
It is proven in \cite{Cecotti:2015lab} that this formula agrees with the vacuum character of the Virasoro minimal model $M(2, 2n+3)$. 

It has been suspected that an analogous IR formula should exist for the Macdonald refinement. This is partly motivated by the observation that $E_q(X)$ is identical to the Schur index of a free hypermultiplet. Therefore, one can consider replacing $E_q(X)$ by $E_{q, t}(X)$ given as \cite{Song:2016yfd} 
\begin{align}
 E_{q, t}(X) = \prod_{i\ge 0} (1- t^\frac{1}{2} X q^i)^{-1} = \sum_{m \ge 0} \frac{t^\frac{m}{2} X^m}{(q)_m} \ , 
\end{align}
which indeed computes the Macdonald index of the free hypermultiplet theory. Empirically, we find that the following ad hoc modification computes the Macdonald refinement: 
\begin{align}
    I_{\textrm{Mac}} = (q)_\infty^r (t)_\infty^r ~ \tr \left[ \prod_{I \textrm{~sources}} E_{q}({X}_{\gamma_I}) \prod_{J \textrm{~sinks}}  E_{q, t}(X_{\gamma_J}) \prod_{I \textrm{~sources}} \hat E_{q}({X}_{-\gamma_I}) \prod_{J \textrm{~sinks}}  E_{q, t}(X_{-\gamma_J}) \right] ,
\end{align}
where $(t)_n$ denotes $(t;q)_n$.
Here we replaced $E_q(X_\gamma)$ by $E_{q, t}(X_\gamma)$ for the sink nodes, and replaced $E_q(X_\gamma)$ by $\hat E_q({X}_\gamma)$ for the anti-particles in the source nodes, where 
\be
\hat E_{q}(X) = \sum_{m\geq 0} \frac{(q^{\frac12}X)^m}{(t;q)_m}\ .
\ee
We experimentally find that this `refined IR formula' applies to a large class of $(A_1,G)$ type theories in the canonical chamber. 
Let us emphasize that this prescription is well-defined only in this particular chamber. 
While we do not have a direct justification for this formula, we will give a physical argument for the formula based on the three-dimensional twisted reduction of 4d theory in the following sections. 

\paragraph{$(A_1, A_{2n})$ theory}
For the $(A_1, A_{2n})$ Argyres-Douglas theory, the proposed refinement procedure gives the following conjectural formula for the Macdonald index:
\begin{equation} \label{eq:A1A2kMacIR}
\boxed{
    I_{\textrm{Mac}}^{A_{2n}}(q, t) = (q)_\infty^n (t)_\infty^n \sum_{m_1, \ldots, m_{2n} \ge 0} \frac{q^{\sum_{i=1}^{2n} m_i + \sum_{i=1}^{2n-1} m_i m_{i+1} }}{\prod_{i \textrm{~even}} (q)_{m_i}^2 \prod_{j \textrm{~odd}}(q)_{m_j} (t)_{m_j}} T^{\sum_{i=1}^n m_{2i}}\ ,
    }
\end{equation}
where $t=qT$. We find that this formula agrees with the Macdonald index computed via other methods to very high orders in $q$.
If we compare this with the Schur index, the prefactor that comes from $n$ copies of abelian vector multiplets $(q)_\infty^{2n}$ is replaced by that of the Macdonald index $(q)_\infty^n (t)_\infty^n$, and the Macdonald grading fugacity $T$ is introduced for the `sink nodes' ($n_2, n_4, \cdots$). In addition, the factor of $(q)_m^2$ in the denominator is replaced by $(q)_m (t)_m$ for the `source nodes' ($n_1, n_3, \cdots$).

\paragraph{$(A_1, D_{2n+1})$ theory}
Following the same procedure, we conjecture that the Macdonald index for the $(A_1, D_{2n+1})$ Argyres-Douglas theory is given by the following expression:
\begin{equation}
\boxed{
\begin{split}
    &I_{\text{Mac}}^{D_{2n+1}}(q,t,z)=
	(q)_{\infty }^{n}(t)_\infty^n \\
	&\qquad \times \sum _{k,\ell \in \mathbb{Z} ^{2n+1}}
	\frac{
	q^{\frac{1}{2}k^{t}b^{D_{2n+1}}\ell}
	q^{\frac12\sum _{i}^{2n+1}(k_{i}+\ell_{i})}z^{\ell_{2n+1}-k_{2n+1}}
	T^{\frac{1}{2}(\sum _{i=1}^{n}(k_{2i}+\ell_{2i})+k_{2n+1}+\ell_{2n+1})}}
	{(q)_{k_{2n+1}}(q)_{\ell_{2n+1}}\prod_{i=1}^{n}(t)_{k_{2i-1}}(q)_{\ell_{2i-1}}\prod _{j=1}^{n}(q)_{k_{2j}}(q)_{\ell_{2j}}}\\
	&\qquad \qquad \times \left(\prod_{i=1}^{2n-1}\delta_{k_{i},\ell_{i}}\right)\delta_{k_{2n}+k_{2n+1},\ell_{2n}+\ell_{2n+1}}
\end{split}
}
\end{equation}
Here $z$ is the $SU(2)$ flavor fugacity. We find that this expression agrees with the Macdonald index computed via other methods to high orders in $q$. 

\paragraph{$(A_1, E_n)$ theory}
Likewise, we obtain the following formula for the $(A_1, E_6)$ theory as
\begin{equation}
\boxed{
    I_{\textrm{Mac}}^{E_6}(q, t)=(q)_\infty^3 (t)_\infty^3\sum_{m\in \mathbb{Z}^6}
    \frac{q^{\frac{1}{2}\sum_{i,j=1}^6b_{ij}^{E_6}}q^{\frac12\sum_{i=1}^6 2m_i}T^{m_2+m_4+m_6}}{\left(\prod_{i=1,3,5}(t)_{m_i}(q)_{m_i}\right)\left(\prod_{i=2,4,6}(q)_{m_i}^2\right)}
} \ . 
\end{equation}
For the $(A_1, E_7)$ theory, we have
\begin{equation}
\boxed{
    I_{\textrm{Mac}}^{E_7} =(q)_\infty^3 (t)_\infty^3\sum_{m\in\mathbb{Z}^8}\frac{q^{\frac{1}{2} m^T b^{E_7} m} q^{\frac12\sum_{i=1}^62m_i+m_7+m_8}T^{m_2+m_4+m_6+\frac{1}{2}(m_7+m_8)}z^{m_8-m_7}}{\left(\prod_{i=1}^8(q)_{m_i}\right)\left(\prod_{j=1,3,5}(t)_{m_j}\right)(q)_{m_2}(q)_{m_4+m_7-m_8}(q)_{m_6-m_7+m_8}}
} \ , 
\end{equation}
and for the $(A_1, E_8)$ theory, we have
\begin{equation}
    \boxed{
     I_{\textrm{Mac}}^{E_8}(q, t) =(q)_\infty^4 (t)_\infty^4\sum_{m\in \mathbb{Z}^8}
    \frac{q^{\frac{1}{2}\sum_{i,j=1}^8b_{ij}^{E_8}m_im_j}q^{\frac12\sum_{i=1}^8 2m_i}T^{m_2+m_4+m_6+m_8}}{\left(\prod_{i=1,3,5,7}(t)_{m_i}(q)_{m_i}\right)\left(\prod_{i=2,4,6,8}(q)_{m_i}^2\right)}
    } \ . 
\end{equation}

\paragraph{$(A_1, A_{2n+1})$ theory}
We propose the Macdonald index for the $(A_1, A_{2n+1})$ Argyres-Douglas theory following the same prescription, but here the roles of the sinks and sources are interchanged
\begin{align}\label{eq:traceMacA2n1}
    I_{\textrm{Mac}} = (q)_\infty^n (t)_\infty^n ~ \tr \left[ \prod_{I \textrm{~sources}} E_{q, t}(X_{\gamma_I}) \prod_{J \textrm{~sinks}} E_{q}({X}_{\gamma_J}) \prod_{I \textrm{~sources}} E_{q, t}(X_{-\gamma_I}) \prod_{J \textrm{~sinks}} \hat E_{q}({X}_{-\gamma_J}) \right] . 
\end{align}
This gives the following expression for the Macdonald index
\begin{equation}
\boxed{
\begin{split}
	I_{\text{Mac}}^{A_{2n+1}}(q,t,z)= (q)_{\infty }^{n}(t)_\infty^n \sum_{k,\ell \in \mathbb{Z}_{\ge 0}^{2n+1}}& \frac{q^{\frac{1}{2}k^{t}b^{A_{2n+1}}\ell}q^{\frac12\sum _{i}^{2n+1}(k_{i}+\ell_{i})}z^{\ell_{1}-k_{1}}T^{\frac{1}{2}\sum_{i=1}^{n+1}(k_{2i-1}+\ell_{2i-1})}}{\prod_{i}^{n+1}(q)_{k_{2i-1}}(q)_{\ell_{2i-1}}\prod _{j=1}^{n}(t)_{k_{2j}}(q)_{\ell_{2j}}}\\
	&\times  \prod_{i=1}^{n}\delta_{(-1)^{i+1}k_{1}+k_{2i+1},(-1)^{i+1}\ell_{1}+\ell_{2i+1}}\prod_{j=1}^{n}\delta_{k_{2j},\ell_{2j}}
\end{split}
}
\end{equation}
where $b^G$ is defined to be $b^G = 2I - C[G]$ where $I$ is the identity matrix, and $C[G]$ is the Cartan matrix of the Lie algebra $G$. This theory has $U(1)$ flavor symmetry ($SU(2)$ for $n=1$), and $z$ is the corresponding flavor fugacity. We have checked that this expression agrees with the known ones to high orders in $q$. 

\paragraph{$(A_1, D_{2n+2})$ theory}
Similarly, with \eqref{eq:traceMacA2n1}, we find the following expression for the Macdonald index of $(A_1, D_{2n+2})$ theory: 
\begin{equation}
\boxed{
\begin{split}
	&I_{\text{Mac}}^{D_{2n+2}}(q,t,z)= (q)_{\infty }^{n}(t)_\infty^n \\
	&~~ \times 
	\sum _{k,\ell \in \mathbb{Z} ^{2n+2}} 
	\frac{
	q^{\frac{1}{2}k^{t}b^{D_{2n+2}}\ell}
	q^{\frac12\sum _{i}^{2n+2}(k_{i}+\ell_{i})}
	z_1^{\ell_{2n+2}-k_{2n+2}}z_2^{\ell_{1}-k_{1}}
	T^{\frac{1}{2} \sum _{i=1}^{n+1}(k_{2i-1}+\ell_{2i-1})+k_{2n+2}+\ell_{2n+2}}}
	{(q)_{k_{2n+1}}(q)_{k_{2n+2}}(q)_{\ell_{2n+1}}(q)_{\ell_{2n+2}}\prod_{i=1}^{n}(q)_{k_{2i-1}}(q)_{\ell_{2i-1}}\prod _{j=1}^{n}(t)_{k_{2j}}(q)_{\ell_{2j}}}\\
	&~~\times
	\left(\prod_{i=1}^{n}\delta_{k_{2i},\ell_{2i}}\right)
	\left(\prod _{j=1}^{n-1}\delta_{(-1)^{j+1}k_{1}+k_{2j+1},(-1)^{j+1}\ell_{1}+\ell_{2j+1}}\right)
	 \delta_{(-1)^{n+1}k_{1}+k_{2n+1}+k_{2n+2},(-1)^{n+1}\times \ell_{1}+\ell_{2n+1}+\ell_{2n+2}}.
\end{split}
}
\end{equation} 
Here $z_1, z_2$ are the flavor fugacities of $SU(2) \times U(1)$ flavor symmetry. Here, $z_1$ and $z_2$ correspond to some linear combinations of the Cartans. The precise map will be given in Section \ref{subsec:Deven}. We have checked that this expression agrees with known ones to high orders in $q$.

\section{Macdonald index from the twisted compactification}

In this section, we describe our strategy to justify the fermionic sum formula for the Macdonald index proposed in the previous section via twisted dimensional reduction to a 3d theory.  
The procedure is as follows:
\begin{enumerate}
    \item Compute the Schur index from the IR trace formula $I(q) = (q)_\infty^{2r} ~\tr \mathcal{O}(q)$, which can be thought of as the half-index of a 3d theory upon $U(1)_r$-twisted reduction. 
    \item Extract the (matter contents and their charges of) 3d $\mathcal{N}=2$ Abelian Chern-Simons matter theory from the above expression for the half-index. 
    \item In general, this gauge theory admits a superpotential deformation, which triggers an RG flow to $\mathcal{N}=4$ SCFT in the IR. The superpotential is expected to break most of the global symmetry except for the distinguished $U(1)_A$ symmetry (and Higgs branch flavor symmetries of 4d), which becomes part of the $SO(4)_R$ symmetry in the IR. The task is to identify this $U(1)_A$ symmetry. 
    \item Refine the half-index by turning on the fugacity $T$ for $U(1)_A$. This gives the Macdonald index of the 4d theory.
\end{enumerate}
The most non-trivial part of the procedure is identifying the distinguished $U(1)_A$ symmetry in the 3d gauge theory. We will describe our method in this section.

\paragraph{3d Abelian Chern-Simons theory from the half-index} The starting point of our procedure is an observation that the fermionic sum formula from the IR formula can be viewed as a twisted half-index of a 3d $\mathcal{N}=2$ Abelian Chern-Simons matter theory (ACSM) \cite{Gaiotto:2024ioj}.
The half-index of a 3d $\mathcal{N}=2$ theory with a supersymmetric boundary condition $B$ is defined by \cite{Gadde:2013wq,Gadde:2013sca,Yoshida:2014ssa,Dimofte:2017} 
\begin{align}
	\II(q)=\mathrm{tr}_B (-1)^{F}q^{J+\frac{R}{2}},
\end{align} 
where the trace is taken over the space of boundary local operators.

Consider a $U(1)^{N}$ Chern-Simons theory with level matrix $K$ and $M$ chiral multiplets $\Phi_{\alpha}$ with $U(1)_{i}$ gauge charge $Q_{i\alpha}$, and $R$-charge  $R_{\alpha}$.
Imposing the supersymmetric Dirichlet boundary conditions to all the vector and chiral multiplets, we obtain the half-index \cite{Dimofte:2017} 
\begin{align}
\II(q,{\boldsymbol x}) = \frac{1}{(q;q)_\infty^N} \sum_{\mathbf{m}\in \mathbb{Z} ^{N}} q^{\frac{1}{2}\mathbf{m}^{t}K\mathbf{m} + \frac12 \boldsymbol{\rho}\cdot \mathbf{m}} {\boldsymbol x}^{K{\bf m}}
	 \prod _{\alpha=1}^{M}\left(q^{R_{\alpha}/2} q^{1-{\mathbf m}\cdot \boldsymbol{Q}_\alpha} {\boldsymbol x}^{-\boldsymbol{Q}_\alpha};q\right)_{\infty }\ ,
\end{align}
where $\rho_{i}$ denotes the mixed Chern-Simons level between the gauge group $U(1)_{i}$ and background $U(1)_R$ symmetry, and $x_i$'s are fugacity for the boundary current $U(1)^N_\partial$. Here we use the standard abbreviation for the multiple fugacities for the symmetry as $\boldsymbol{x}^{\boldsymbol{m}} \equiv \prod_i x_i^{m_i}$. In most of the examples below, we consider the fully unrefined index with ${\boldsymbol x}\rightarrow 1$.

Upon comparing this expression with the one obtained from the IR trace formula for the Schur index \eqref{eq:traceSchur}, we extract an abelian Chern-Simons theory. For example, the Schur index of the form \eqref{eq:A1A2nIR} for the $(A_1, A_{2N})$ theory gives rise to $U(1)^{2N}$ Chern-Simons (CS) theory, where each $U(1)$ is coupled to 2 chiral multiplets of charge $1$, giving a total of $M=4N$ chiral multiplets. We can also read the CS level matrix $K$, mixed $U(1)-R$ CS level $\rho$ for each $U(1)$ gauge factors, and $R$-charges for the chiral multiplets by comparing the formulae. 

This description of the abelian Chern-Simons theory can be motivated by considering a Janus configuration in the Coulomb branch of the 4d SCFT, induced by the $U(1)_r$ twisting. See section 6 of \cite{Gaiotto:2024ioj} for further discussion.

\paragraph{Supersymmetry enhancement}
The abelian Chern-Simons theory obtained in this way is expected to flow to an SCFT with $\CN=4$ supersymmetry enhancement, after a suitable superpotential deformation. The IR $\CN=4$ SCFT has an $\CN=2$ superconformal subalgebra where $U(1)_R$ symmetry embeds into the full $\CN=4$ R-symmetry group $SU(2)_H\times SU(2)_C$, leaving $U(1)_A= U(1)_H-U(1)_C$ as a distinguished global symmetry. When the 3d $\CN=4$ SCFT is expected to have a zero-dimensional Coulomb/Higgs branch, we anticipate that there exists a suitable superpotential deformation that lifts most of the global symmetry of the UV ACSM.
This leaves one combination of $U(1)$, which can be identified with the IR $U(1)_A$ symmetry. To find the superconformal $R$-symmetry, we consider a one-parameter mixing of the $R$-symmetry
\begin{align}
U(1)_{R_{\nu}} = U(1)_R - \nu U(1)_A\ ,
\end{align}
and perform $F$-maximization \cite{Jafferis:2010un}. We define our normalization
so that $\nu=0$ corresponds to the superconformal $R$-symmetry $U(1)_{R_0} = U(1)_R = U(1)_C + U(1)_H$. 

Accordingly, one can introduce a refined half-index, defined as
\begin{align}
\II(q,\nu,T) =\mathrm{tr}_B (-1)^{F}q^{J+\frac{R_\nu}{2}} T^{A}\ .
\end{align}
In the special limit $\nu=-1$, the half-index computes the index with the twisted spin $J + R_\nu/2 = J +R_H$, realizing the topological $A$-twist. We therefore define 
\begin{align}\label{A refined half}
\II^A(q,T) :=\II(q,-1,T)\ , 
\end{align}
which corresponds to the refinement associated with the Macdonald grading of the parent 4d SCFT.

\paragraph{Identifying $U(1)_{A}$} The 3d $\CN=4$ SCFTs obtained by twisted compactification of $(A_1, G)$ Argyres-Douglas theories are expected to have zero-dimensional Coulomb branches, while their Higgs branches coincide with those of the original 4d SCFT. On the other hand, the proposed $\CN=2$ ACSM descriptions generally exhibit a larger global symmetry than the Higgs branch symmetry. We therefore expect that there exists an appropriate superpotential deformation that breaks the global symmetries to $U(1)_A\times T_H$, where $T_H\subset G_H$ denotes the Cartan subgroup of the Higgs branch symmetry. 

The superpotential deformation generally involves disorder chiral primary operators constructed by dressed monopole operators. As a first step, we identify all half-BPS gauge-invariant monopole operators in a given ACSM description, which may serve as candidates for the superpotential deformation.

For low-rank examples, one can identify collections of superpotential deformations that are compatible with the Higgs branch global symmetry, which in turn provides candidates for the $U(1)_A$ symmetry. Then for each such candidate, we perform F-maximization \cite{Jafferis:2010un} with respect to $U(1)_A$ and calculate the superconformal index \cite{Kim:2009wb, Imamura:2011su}
\begin{align}
	\mathcal{I}(\mathfrak{q},T)=\mathrm{tr} _{\mathbb{S}^{2}}(-1)^{F}\mathfrak{q}^{\frac{R}{2}+j_{3}}T^{A}\ ,
\end{align}
where the trace is over the states with $E-R-j_{3}=0$. The resulting index is then tested against non-trivial evidence for the supersymmetry enhancement \cite{Gang:2018huc,Gang:2024loa}:
\begin{itemize}
\item[1.] The superconformal index contains a term $q^{3/2}(T+T^{-1})$, signaling the existence of the extra supercurrent multiplets in the $\CN=4$ algebra\footnote{This condition is neither necessary nor sufficient, since there may exist, for example, chiral primary operators with $R$-charge 3. In fact, we have examples where supersymmetry enhancement is expected, yet this term is partially or entirely absent.}, as well as a possible moment map operator for the Higgs branch symmetry appearing at $q^{1/2}$ order (when such a symmetry is present). 
\item[2.] The Coulomb branch Hilbert series is trivial.\footnote{For a 3d SCFT, the superconformal index can be written as
\begin{align}
	\mathcal{I}(\mathfrak{q},\mathfrak{t})=\mathrm{tr}_{\mathbb{S}^{2}}(-1)^{F}\mathfrak{q}^{j_{3}+\frac{1}{2}(R_{H}+R_{c})}\mathfrak{t}^{R_{H}-R_{C}}
\end{align}
and it gets contribution only from states satisfying $E-R_{H}-R_{C}-j_{3}=0$.
We can rewrite the index by
\begin{align}
	\mathcal{I}(\mathfrak{q},\mathfrak{t})=\mathrm{tr} (-1)^{F}(\mathfrak{q}^{\frac{1}{2}}\mathfrak{t})^{E-R_{c}}(\mathfrak{q}^{\frac{1}{2}}\mathfrak{t}^{-1})^{E-R_{H}}.
\end{align} 
The Higgs and Coulomb branch Hilbert series are obtained by taking the limit \cite{Razamat:2014}
\begin{subequations}
\begin{align}
    \text{Coulomb:}&\qquad \mathfrak{q}\rightarrow 0,\mathfrak{t}\rightarrow 0,\qquad \mathfrak{q}^{\frac{1}{2}}\mathfrak{t}^{-1}\;\label{eq:coulomb}\text{fixed}\\
	\text{Higgs:}&\qquad \mathfrak{q}\rightarrow 0,\mathfrak{t}\rightarrow \infty ,\qquad \mathfrak{q}^{\frac{1}{2}}\mathfrak{t}\;\text{fixed}\label{eq:higgs}
\end{align}
\end{subequations}
which give extra shortening condition $E=R_{H}$ and $E=R_{C}$ for the trace, respectively.
}
\item[3.] The Higgs branch Hilbert series agrees with that of the Higgs branch chiral ring of 4d SCFT.
\end{itemize}

In some low-rank examples, this procedure determines the $U(1)_A$ symmetry in the ACSM description, and one can check that the refined half-index \eqref{A refined half} reproduces the conjectural Macdonald index formula described in section \ref{sec: conjectural Macdonald}.  

One subtlety in identifying $U(1)_A$ arises from the fact that a global symmetry can be redefined by mixing it with gauge symmetries.
The Dirichlet half-index is sensitive to this choice because, at the boundary, the gauge symmetry is reduced to a global symmetry. Therefore, the computation of the refined half-index is not completely canonical and depends on this choice. In the examples below, we show that there exists a representation of the $U(1)_A$ symmetry that reproduces the conjectural Macdonald index presented in section \ref{sec: conjectural Macdonald}, providing a strong consistency check despite its non-uniqueness.

For higher rank examples, where F-maximization is not practically feasible, we begin with the assumption that the Schur index coincides with the A-twisted half-index of the 3d $\CN=4$ SCFT, and that the conjectural Macdonald index introduced in section \ref{sec: conjectural Macdonald} computes its refined half-index. This assumption completely determines the data for the superconformal R-symmetry. We then compute the corresponding superconformal index and the Coulomb and Higgs branch Hilbert series, and test them against the conditions described above. 

\section{Examples}
In this section, we apply our procedure for $(A_1, G)$ type Argyres-Douglas theories to recover our conjectural closed-form expression for the Macdonald index. 

\subsection{$(A_1, A_{2n})$ theory}
\begin{figure}
  \centering
  \begin{tikzpicture}
			\node[W] (1) at (0,0){};
		 	\node[above=3mm] at (1) {$\gamma_1$};
            
            \node[W] (2) at (2,0){};
		 	\node[above=3mm] at (2) {$\gamma_2$};

            \node[] (3) at (4,0){$\cdots$};
            
            \node[W] (4) at (6,0){};
		 	\node[above=3mm] at (4) {$\gamma_{2n}$};
            
			\draw[->] (1)--(2);
            \draw[<-] (2)--(3);
            \draw[->] (3)--(4);
  \end{tikzpicture}
  \caption{\label{fig:a2nquiver}
  BPS quiver for the $(A_1, A_{2n})$ theory in the canonical chamber}
\end{figure}
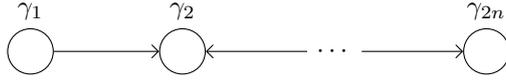

The BPS quiver for the $(A_1, A_{2n})$ theory is given in Figure~\ref{fig:a2nquiver}.
This describes the canonical chamber with an equal number of sources and sinks. The trace formula \eqref{eq:traceSchur} gives the Schur index \cite{Cordova:2015nma},
\begin{align}\label{eq:a2nSchur}
	I^{A_{2n}}_{\text{Schur}}(q)=(q)_{\infty }^{2n}\sum _{k\in \mathbb{Z}_{\ge 0} ^{2n}} \frac{q^{\frac{1}{2}k^{t}b^{A_{2n}}k}q^{\sum _{i}k_{i}}}{\prod _{i=1}^{2n}(q)_{k_{i}}^2}
    = \frac{1}{(q)_\infty^{2n}} \sum_{k \in \mathbb{Z}^{2n}} q^{\frac{1}{2}k^{t}b^{A_{2n}}k + \sum_i k_i } \prod_{i=1}^{2n} (q^{1+k_i})_\infty^2 \ ,
\end{align}
where $b^{g}_{ij}=2\delta_{ij}-C[g]_{ij}$ and $C[g]$ is the Cartan matrix of $g$. This expression agrees with the character of the Virasoro minimal model $M(2, 2n+3)$ \cite{Cecotti:2015lab}. 

We begin with the observation that this formula coincides with the Dirichlet half-index of 3d $U(1)^{2n}$ Chern-Simons theory with level
\begin{align}
	K=b^{A_{2n}} 
     = \begin{pmatrix}
		0 & 1 & 0 & 0 & \cdots & 0 & 0\\
        1 & 0 & 0 & 0 & \cdots & 0 & 0\\
        0 & 0 & 0 & 1 & \cdots & 0 & 0\\
        0 & 0 & 1 & 0 & \cdots & 0 & 0\\
        &&& \vdots \\
        0 & 0 & 0 & 0 & \cdots & 0 & 1 \\
        0 & 0 & 0 & 0 & \cdots & 1 & 0
	\end{pmatrix}\ .
\end{align} 
Additionally, we have two chiral multiplets $\Phi_i$ and $\tilde{\Phi}_i$ with charge $-1$ for each $U(1)_{i} \subset U(1)^{2n}$, giving a total of $4n$ chiral multiplets. Therefore, we can write the matter content and the charge matrix $Q_{i\alpha}$ as 
\begin{align}
    \left[\begin{array}{c|c} & \text{matters} \\ \hline \text{gauge group} & Q \end{array}  \right] 
    = \left[ \begin{array}{c|ccccccc}
      & \Phi_1 & \tilde\Phi_1 & \Phi_2 & \tilde\Phi_2 & \cdots & \Phi_{2n} & \tilde\Phi_{2n}\\
      \hline
      U(1)_1 & -1 & -1 & 0 & 0 &\cdots & 0 & 0 \\
      U(1)_2 & 0 & 0 & -1 & -1 & \cdots & 0 & 0 \\
       & &&&\vdots \\
      U(1)_{2n} & 0 & 0 & 0 & 0 & \cdots & -1 & -1 
    \end{array} \right] \ . 
\end{align}
The half-index formula shows that the $R$-charges for all chiral multiplets are zero and the mixed Chern-Simons level between $U(1)_R$ and $U(1)_i$ is $\rho_{i}=2$ for all $i=1,2,\cdots,2n$. 
Before adding the superpotential term, the ACSM theory has $U(1)^{4n}$ flavor symmetry. 

We expect that there exists a superpotential deformation that breaks all the symmetries except for one $U(1)$, which we denote by $U(1)_A$. Our task is to identify this symmetry. We further expect that the Schur index \eqref{eq:a2nSchur}, when interpreted as the half-index of the 3d $\CN=4$ theory, is computed in the A-twisted limit. 

\subsubsection{$A_{2}$}
Let us start with the simplest case of $n=1$. The trace formula gives
\begin{align}
	I_{\text{Schur}}^{A_{2}}(q)=(q)_{\infty }^2\sum _{k_{1},k_{2} \ge 0} \frac{q^{k_{1}k_{2}+k_{1}+k_{2}}}{(q)_{k_{1}}^2(q)_{k_{2}}^2}
    = \frac{1}{(q)_\infty^2} \sum_{k_1, k_2 \in \mathbb{Z}} q^{k_1 k_2 + k_1 + k_2} (q^{1+k_1})_\infty^2 (q^{1+k_2})_\infty^2 \ . 
\end{align}
This expression can be identified with the half-index of the 3d $\mathcal{N}=2$ Abelian Chern-Simons theory with $U(1)^2$ gauge group, with the following data:
\begin{align}
    \begin{split}
        K = 
        \begin{pmatrix}
		0&1\\1&0
	    \end{pmatrix} \ , 
	       \quad Q = 
           \begin{pmatrix}
           -1 & -1 & 0 & 0 \\
           0 & 0 & -1 & -1 
	        \end{pmatrix} \ , \quad
            \rho = \begin{pmatrix} 2 & 2 \end{pmatrix}
    \end{split}
\end{align}
Here $K$, $Q$, and $\rho$ refer to the Chern-Simons level, charge matrix, and $R$-gauge mixed Chern-Simons level, respectively. All the chiral multiplets have $R$-charge zero.

This gauge theory has $U(1)^4$ flavor symmetry, corresponding to the rotations of each chiral multiplet. We expect that there exists an appropriate superpotential deformation constructed from dressed monopole operators which break $U(1)^4$ flavor symmetry down to the single $U(1)_A$. Such monopole operators should have $R$-charge 2, and there are four of them in total, as listed in Table~\ref{tab:a2Mono}. 
\begin{table}[t]
\centering
\begin{tabular}{@{}ccc|cccc@{}}
\toprule
 & $m_1$ & $m_2$ & $\Phi_1$ & $\tilde{\Phi}_1$ & $\Phi_2$ & $\tilde{\Phi}_2$ \\ \midrule
$\surd$      & 1     & 0     &      0    &      0    & 1        & 0         \\
$\surd$      & 1     & 0     &     0     &      0    &      0    & 1        \\
$\surd$      & 0     & 1     & 1        &   0       &    0      &  0        \\
       & 0     & 1     &     0     & 1        &    0      &       0   \\ \bottomrule
\end{tabular}
\caption{Monopole operators of $R=2$ for the 3d theory of $(A_{1},A_{2})$. Each row indicates the dressed monopole operator with magnetic charge $m$ dressed by each chiral multiplet by the power in the corresponding column. Operators included in the superpotential are indicated with $\surd$ in the first column.}
\label{tab:a2Mono}
\end{table}
Consider the following superpotential\footnote{Any choice of three out of four monopole operators gives rises the same result, thanks to the symmetry.}
\begin{align}
W = V_{\{1,0\}}\Phi_{2}+V_{\{1,0\}}\tilde{\Phi}_{2}+V_{\{0,1\}}\Phi_{1} \ .
\end{align} 
The resulting unbroken global symmetry is $U(1)_{\tilde{\Phi}_{1}}$, which rotates $\tilde{\Phi}_{1}$ with charge $+1$ and keeps the rest intact. By mixing with gauge symmetry, this global symmetry can also be represented as a combination of the topological symmetry for the second gauge node and the symmetry that rotates $\Phi_1$ with charge $-1$.\footnote{Consider the bosonic action $2A_1 dA_2 + |D_{(-A_1+A^{\text{bg}})}\tilde\Phi|^2 + |D_{(-A_1)} \Phi|^2+\cdots$. By redefining the gauge connection $A_1\rightarrow A_1+ A^{\text{bg}}$, we obtain $2A_1 dA_2 +2 A^{(\text{bg})} dA_2 + |D_{(-A_1)}\tilde\Phi|^2 + |D_{(-A_1-A^{(bg)})} \Phi|^2+\cdots$}
We will now argue that this is the desired superpotential and verify that this unbroken symmetry is the axial symmetry $U(1)_{A}$ that accounts for the Macdonald grading. 

\begin{figure}[htpb]
	\centering
	\includegraphics[width=0.65\textwidth]{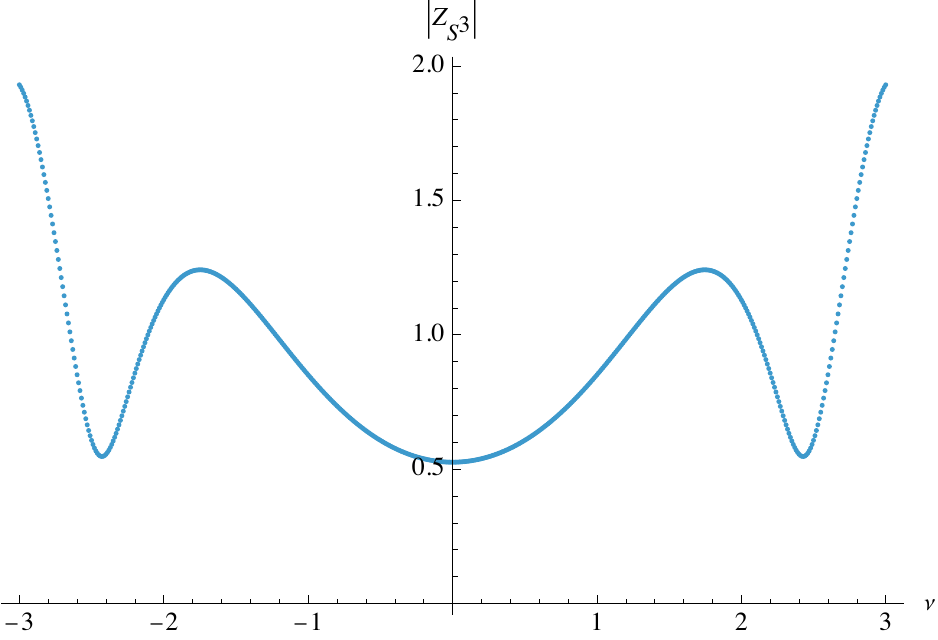}
	\caption{The absolute value of the three-sphere partition function of the 3d theory for $(A_{1},A_{2})$ as a function of mixing parameter $\nu$. The function is minimized at $\nu=0$.}
	\label{fig:a2Fmax}
\end{figure}

We can now compute the superconformal index of the 3d SCFT via the gauge theory description, refined to include the candidate $U(1)_A$ fugacity. In order to determine the superconformal R-charge, we perform F-maximization,via the three-sphere partition function using the Bethe ansatz method of \cite{Closset:2017}.
From the 3d gauge theory, we construct the effective twisted superpotential $W$ as a function of the mixing parameter $\nu$.
The Bethe ansatz equation $\{\exp(2\pi i\partial W(u)/\partial u_{i})=1\}$ for this theory turns out to have two solutions $u_{0}$ and $u_{1}$.
The three-sphere partition function is then given by
\begin{align}
	Z_{S^{3}}=\sum _{u_{0},u_{1}}\mathcal{H}^{-1}(u)\mathcal{F}^{1}(u).
\end{align} 
We plot the $S^3$ partition function as a function of mixing parameter $\nu$ in Figure \ref{fig:a2Fmax}. We see that the absolute value $|Z_{S^{3}}|$ is minimized (or $F$ is maximized) at $\nu=0$, which corresponds to the superconformal point.

The superconformal index at the fixed point is
\begin{align}
	\mathcal{I}_{SCI}(\mathfrak{q},T)=1-\mathfrak{q}+\left(T^{-1}+T\right)\mathfrak{q}^{\frac{3}{2}}-2\mathfrak{q}^2+\left( T^{-1} +T \right)\mathfrak{q}^{\frac{5}{2}}+\cdots.
\end{align}
The presence of the term $(T+T^{-1}) \mathfrak{q}^{\frac{3}{2}}$ provides a strong evidence for the $\mathcal{N}=4$ supersymmetry enhancement.
Now the Coulomb limit is simply obtained by taking the limit $T \to \mathfrak{q}^{1/2}$. For the Higgs limit, we take the limit $\lim_{\mathfrak{q} \to 0}\mathcal{I}_{SCI}(\mathfrak{q},T)|_{T\rightarrow \mathfrak{q}^{-\frac{1}{2}}T} $. We find 
\begin{align}
	\mathcal{I}_{\mathrm{C}}(\mathfrak{q})=1, \quad
	\mathcal{I}_{\mathrm{H}}(T)=1, 
\end{align}
which is consistent with the expectation that both the Coulomb and Higgs branches of the 3d SCFT are trivial. This strongly supports our identification of $U(1)_A$. 

Now, armed with $U(1)_A$ symmetry, we can compute the $T$-refined half-index, in the A-twisted limit. This gives
\begin{align}
\begin{split}
	\II(q,T)&=\frac{1}{(q)_\infty^{2}} \sum_{k_{1},k_{2}\in \mathbb{Z} } q^{k_{1}k_{2}+k_{1}+k_{2}}T^{k_{2}}(q^{1+k_{1}}T)_\infty(q^{1+k_{1}})_\infty(q^{1+k_{2}})_\infty^{2}\\
	&= (q)_\infty(qT)_\infty \sum _{k_{1},k_{2}\in \mathbb{Z}_{\ge 0} } \frac{q^{k_{1}k_{2}+k_{1}+k_{2}}T^{k_{2}}}{(qT)_{k_{1}}(q)_{k_{1}}(q)_{k_{2}}^2}\label{eq:a2Mac}\\
	&= 1+q^2 T+q^3 T+q^4 T+q^5 T+q^6 \left(T^2+T\right)+q^7 \left(T^2+T\right) +q^8 \left(2 T^2+T\right)
    +\cdots \ , 
\end{split}
\end{align}
where we see that the expression in the second line reproduces with our proposal \eqref{eq:A1A2kMacIR}. 
This expression can be shown to agree with the fermionic sum formula \eqref{eq:A1A2nFermion} for the Macdonald index for this case \cite{Gaiotto:2024ioj}. 

\subsubsection{$A_{4}$}
The corresponding 3d $\CN=2$ theory can be read off from the Schur index as before; it is given by a $U(1)^4$ ACSM theory with eight chiral multiplets. We now search for a superpotential deformation given by dressed monopole operators that breaks all the $U(1)^{8}$ flavor symmetries except for $U(1)_A$ and triggers an RG flow to the $\CN=4$ SCFT in the IR.

\begin{table}[t]
\centering
\begin{tabular}{@{}c cccc|ccccccccc@{}}
\toprule
& $m_1$ & $m_2$ & $m_3$ & $m_4$ & $\Phi_1$ & $\tilde{\Phi}_1$ & $\Phi_2$ & $\tilde{\Phi}_{2}$ & $\Phi_3$ & $\tilde{\Phi}_3$ & $\Phi_4$ & $\tilde{\Phi}_4$ \\
\midrule
$\surd$           & 0 & 0 & 0 & 1 & 0 & 0 & 0 & 0 & 0 & 1 & 0 & 0 \\
              & 0 & 0 & 0 & 1 & 0 & 0 & 0 & 0 & 1 & 0 & 0 & 0 \\
$\surd$          & 0 & 0 & 1 & 0 & 0 & 0 & 0 & 1 & 0 & 0 & 0 & 1 \\
$\surd$           & 0 & 0 & 1 & 0 & 0 & 0 & 0 & 1 & 0 & 0 & 1 & 0 \\
$\surd$          & 0 & 0 & 1 & 0 & 0 & 0 & 1 & 0 & 0 & 0 & 0 & 1 \\
              & 0 & 0 & 1 & 0 & 0 & 0 & 1 & 0 & 0 & 0 & 1 & 0 \\
$\surd$          & 0 & 1 & 0 & 0 & 0 & 1 & 0 & 0 & 0 & 1 & 0 & 0 \\
             & 0 & 1 & 0 & 0 & 0 & 1 & 0 & 0 & 1 & 0 & 0 & 0 \\
              & 0 & 1 & 0 & 0 & 1 & 0 & 0 & 0 & 0 & 1 & 0 & 0 \\
$\surd$         & 0 & 1 & 0 & 0 & 1 & 0 & 0 & 0 & 1 & 0 & 0 & 0 \\
              & 1 & 0 & 0 & 0 & 0 & 0 & 0 & 1 & 0 & 0 & 0 & 0 \\
$\surd$       & 1 & 0 & 0 & 0 & 0 & 0 & 1 & 0 & 0 & 0 & 0 & 0 \\
\bottomrule
\end{tabular}
\caption{Monopole operators in the 3d theory for $(A_{1},A_{4})$. Each row indicates the dressed monopole operator with magnetic charge $m$, dressed by each chiral multiplets by the power in the corresponding column. Operators included in the superpotential is indicated with a check mark. }
\label{tab:a4Mono}
\end{table}
There are 12 candidate gauge invariant half-BPS dressed monopole operators which can appear in the superpotential, listed in Table~\ref{tab:a4Mono}.
There are 22 possible candidates for $U(1)_{A}$ left unbroken by turning on 7 independent monopole operators.
By requiring that the Coulomb and Higgs limits of the superconformal index are trivial, we are left with only two cases (up to the symmetry $U(1)_{1}\leftrightarrow U(1)_{4},U(1)_{2}\leftrightarrow U(1)_{3}$ and $\Phi\leftrightarrow \tilde{\Phi}$):
\begin{align}
	-U(1)_{\Phi_{1}}+U(1)_{\Phi_{3}},\quad U(1)_{\Phi_{1}}+U(1)_{\tilde{\Phi}_{1}}-U(1)_{\Phi_{3}}.
\end{align}

\begin{figure}[htpb]
	\centering
	\includegraphics[width=0.65\textwidth]{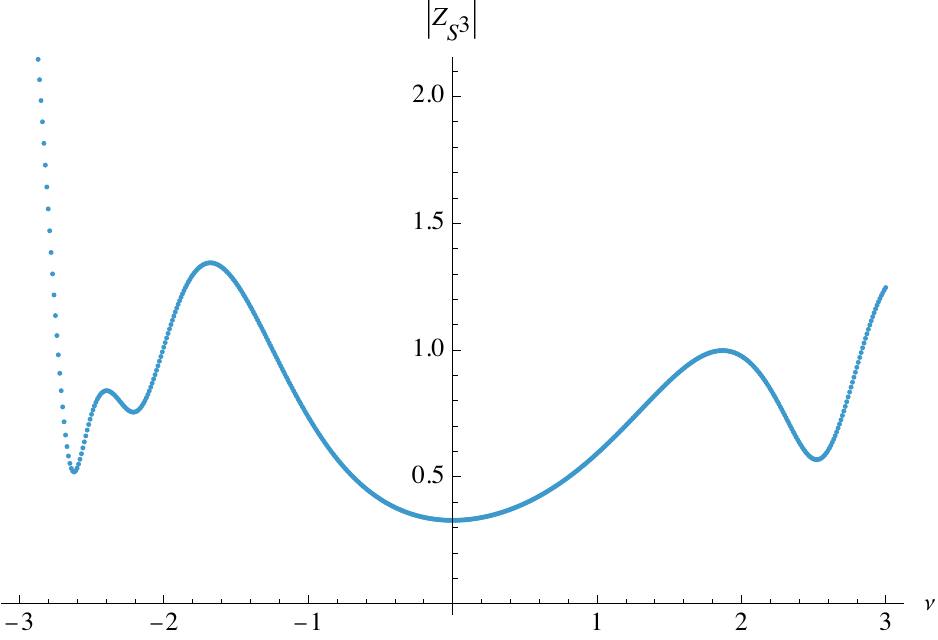}
	\caption{The absolute value of three-sphere partition function for the 3d theory of $(A_{1},A_{4})$ as a function of the mixing parameter $\nu$ with $U(1)_{A}=-U(1)_{\Phi_{1}}+U(1)_{\Phi_{3}}$. The function is minimized at $\nu=0$.}
	\label{fig:a4Fmax}
\end{figure}
Among them, only for the first choice, we find that the $F$ is maximized at $\nu=0$ (see Figure~\ref{fig:a4Fmax}) and the superconformal index at the fixed point gives
\be 
	\mathcal{I}_{SCI}(\mathfrak{q},T)=1-\mathfrak{q}+\left(T^{-1}+T\right)\mathfrak{q}^{\frac{3}{2}}-2\mathfrak{q}^2+T^{-1} \mathfrak{q}^{\frac{5}{2}}+\cdots .
\ee
which agrees with the expectation that the supersymmetry is enhanced to $\CN=4$ at the fixed point. One can also check that
\be
    \mathcal{I}_{C}(\mathfrak{q})=1,\qquad
	\mathcal{I}_{H}(T)=1.
    \ee
as expected.

Hence we conclude
\begin{align}
	U(1)_{A}=-U(1)_{\Phi_{1}}+U(1)_{\Phi_{3}}\ .
\end{align} 
This symmetry can be alternatively realized as
\begin{align}
U(1)_A = U(1)_{T_2} + U(1)_{T_4} - U(1)_{\Phi_1} - U(1)_{\tilde\Phi_3} \ , 
\end{align}
by mixing with the gauge symmetry for the third gauge node, $U(1)_3$.

The $T$-refined half-index in the A-twisted limit reproduces the conjectural Macdonald index of $(A_{1},A_{4})$ AD theory
\begin{align} \label{eq:a4Mac}
\begin{split}
		\II(q,T)&=\frac{1}{(q)_\infty^{4}}\sum _{k\in\mathbb{Z} ^{4}}q^{\frac{1}{2}k^{t}b^{A_{4}}k+\sum _{i=1}^{4}k_{i}}T^{k_{2}+k_{4}} 
        \prod_{i=2, 4} (q^{1+k_i})_\infty^2 \prod_{j=1, 3} (q^{1+k_j}T)_\infty (q^{1+k_j})_\infty \\
	&=(q)_\infty^2(qT)_\infty^2\sum _{k\in\mathbb{Z} ^{4}} \frac{q^{\frac{1}{2}k^{t}b^{A_{4}}k+\sum _{i=1}^{4}k_{i}}T^{k_{2}+k_{4}}}{(q)_{k_{2}}^2 (q)_{k_{4}}^2 (qT)_{k_{1}}(q)_{k_{1}} (qT)_{k_{3}}(q)_{k_{3}}} \ , 
\end{split}
\end{align}
which can be expanded in $q$-series as
\begin{align}
\begin{split}
	\II(q,T) &=1+q^2 T+q^3 T+q^4 \left(T^2+T\right)+q^5 \left(T^2+T\right)+q^6 \left(2 T^2+T\right)\\
    &\qquad \qquad +q^7 \left(2 T^2+T\right) +q^8 \left(T^3+3 T^2+T\right)+\cdots. 
\end{split}
\end{align}
It agrees with the known $q$-expansion of the Macdonald index. 

\subsubsection{General Form}
Based on these calculations, we conjecture that the superpotential deformation for general $n$ as
\begin{align}
\begin{split}
W &= \sum_{i=1}^{n-1}V_{2i}\Phi_{2i-1}\Phi_{2i+1}+\sum _{i=1}^{n}V_{2i}\widetilde{\Phi}_{2i-1}\widetilde{\Phi}_{2i+1}\\
	&\quad + \sum_{i=1}^{n}V_{2i-1}\Phi_{2i-2}\widetilde{\Phi}_{2i}+\sum _{i=2}^{n}V_{2i-1}\widetilde{\Phi}_{2i-2}\widetilde{\Phi}_{2i}
	+V_{2n-1}\widetilde{\Phi}_{2n-2}\Phi_{2n} \ .
\end{split}
\end{align}
The axial symmetry survives under this deformation is
\begin{align}
    U(1)_A = \sum_i U(1)_{T_{2i}} - (U(1)_{\Phi_1} + U(1)_{\tilde\Phi_3} + U(1)_{\Phi_5} + U(1)_{\tilde\Phi_7} + \cdots )\ , 
\end{align}
where $U(1)_{T_{2i}}$ denotes the topological symmetries for the $2i$-th gauge node.

The refined half-index with this $U(1)_A$ symmetry leads to the following form of the Macdonald index, which reproduces the conjecture in section \ref{sec: conjectural Macdonald}:
\begin{align}
I_{\textrm{Mac}}^{A_{2n}}(q,T)=(q)_{\infty }^{n}(qT)_\infty^n\sum _{k\in \mathbb{Z} ^{2n}} \frac{q^{\frac{1}{2}k^{t}b^{A_{2n}}k}q^{\sum _{i}k_{i}}T^{\sum _{i=1}^{n}k_{2i}}}{\prod _{i=1}^{n}(q)_{k_{2i-1}}(qT)_{k_{2i-1}}\prod _{j=1}^n(q)_{k_{2j}}^{2}}.
\end{align}

\subsection{$(A_1, A_{2n+1})$ theory}

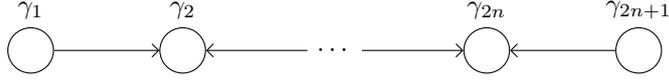
\begin{figure}
  \centering
  \begin{tikzpicture}
			\node[W] (1) at (0,0){};
		 	\node[above=3mm] at (1) {$\gamma_1$};
            
            \node[W] (2) at (2,0){};
		 	\node[above=3mm] at (2) {$\gamma_2$};

            \node[] (3) at (4,0){$\cdots$};
            
            \node[W] (4) at (6,0){};
		 	\node[above=3mm] at (4) {$\gamma_{2n}$};

            \node[W] (5) at (8,0){};
		 	\node[above=3mm] at (5) {$\gamma_{2n+1}$};
            
			\draw[->] (1)--(2);
            \draw[<-] (2)--(3);
            \draw[->] (3)--(4);
            \draw[<-] (4)--(5);
  \end{tikzpicture}
  \caption{\label{fig:a2n1quiver}
  BPS quiver for the $(A_1, A_{2n+1})$ theory in the canonical chamber}
\end{figure}

The BPS quiver of the $(A_{1},A_{2n+1})$ AD theory is given in Figure~\ref{fig:a2n1quiver}. This theory has an $n$-dimensional Coulomb branch, and its Higgs branch is given by $\mathbb{C}^2/\mathbb{Z}_{n+1}$ with $U(1)$ flavor symmetry (for $n\ge2$). 
Here we focus on $n\geq2$ and the
$(A_{1},A_{3}) = (A_{1},D_{3})$ case will be dealt separately when we discuss $(A_1, D_{2n+1})$ theory. 

The $U(1)$ flavor symmetry corresponds to the flavor charge
\begin{align}\label{flavor A odd}
	\gamma_{f}=\sum _{i=0}^{n}(-1)^{i}\gamma_{2i+1},\qquad \mathrm{Tr} [X_{\gamma_{f}}]=z.
\end{align} 

The trace formula \eqref{eq:traceSchur} gives 
\begin{align}\label{eq:a2n1Schur}
	\begin{split}
		I_{\text{Schur}}^{A_{2n+1}}(q,z)=(q)_{\infty }^{2n}\sum _{k,\ell \in \mathbb{Z} ^{2n+1}}& \frac{q^{\frac{1}{2}k^{t}b^{A_{2n+1}}\ell}q^{\frac12\sum _{i=1}^{2n+1}(k_{i}+\ell_{i})}z^{\ell_{1}-k_{1}}}{\prod_{i=1}^{n}(q)_{k_{i}}(q)_{\ell_{i}}}\\
	&\times \prod_{i=1}^{n}\delta_{(-1)^{i+1}k_{1}+k_{2i+1},(-1)^{i+1}\ell_{1}+\ell_{2i+1}}\prod_{j=1}^{n}\delta_{k_{2j},\ell_{2j}}
\end{split}
\end{align}
where $b_{ij}^{g}=2\delta_{ij}-C[g]_{ij}$ and $C[g]$ is the Cartan matrix of $g$.

As before, we start from the observation is that this formula coincides with the half-index of a 3d $U(1)^{2n+2}$ Chern-Simons theory with level
\begin{align}
    K=\left( 
	\begin{array}{ccc|c}
		\multicolumn{3}{c}{\multirow{3}{*}{\raisebox{-1mm}{\scalebox{1.1}{$b^{A_{2n+1}}$}}}}\vrule &0\\
		& & &\vdots\\
		& & & 0\\
		\hline
		0& \cdots  & 0 & 0
	\end{array}
	\right)\ .
\end{align}
The mixed Chern-Simons levels between the gauge $U(1)$s and $R$-symmetry are $\rho=\{2,2,\cdots,2,0\}$ for odd $n$ and $\rho=\{1,2,2,\cdots,2,1\}$ for even $n$.

The Higgs branch flavor symmetry can be idenfied with
\begin{align}
\begin{split}
    U(1)_f=U(1)_{T_1}-U(1)_{T_{2n+1}}
\end{split}
\end{align}
where $U(1)_{T_i}$ denotes the topological symmetry corresponding to the $i$-th gauge $U(1)$.

Also, we have $4n+2$ chiral multiplets, all with trial $R$-charge $0$.
Their gauge charges are
\begin{itemize}
\item $\Phi_i$ has charge $-1$ under $U(1)_i$ for $i=1,2,3,\cdots,2n+2$.
\item $\tilde{\Phi}_j$ has charge $-1$ under $U(1)_j$ for $j=2,4,6,\cdots,2n$.
\item $\hat{\Phi}_k$ has charge $(-1)^\frac{k-1}{2}$ under $U(1)_1$, $-1$ under $U(1)_k$, and $(-1)^\frac{k+1}{2}$ under $U(1)_{2n+2}$ for $k=3,5,7,\cdots,2n+1$.
\end{itemize}

\subsubsection{$A_{5}$}
For $n=2$ the formula \eqref {eq:a2n1Schur} gives 
\begin{align}
\begin{split}\label{schur A5}
	I_{\text{Schur}}^{A_{5}}(q,z) &=(q)_\infty^{4}\sum _{m\in \mathbb{Z} ^{6}} \frac{q^{\frac{1}{2}\sum _{i,j=1}^{5}b^{A_5}_{ij}m_{i}m_{j}}(q^{\frac{1}{2}})^{m_{1}+m_{6}+2(m_{2}+m_{3}+m_{4}+m_{5})}z^{m_{1}-m_{6}}}{\prod _{i=1,3,5,6}(q)_{m_{i}}\prod _{j=2,4}(q)_{m_{j}}^2(q)_{m_{1}+m_{3}-m_{6}}(q)_{-m_{1}+m_{5}+m_{6}}}\ ,
\end{split}
\end{align} 
from which we can extract the $U(1)^{6}$ gauge theory description as described above.

\begin{table}[t]
\centering
\begin{tabular}{@{}c cccccc|cccccccccc@{}}
\toprule
  & $m_1$ & $m_2$ & $m_3$ & $m_4$ & $m_5$ & $m_6$ & $\Phi_1$ & $\Phi_2$ & $\tilde{\Phi}_{2}$ & $\Phi_3$ & $\Phi_{4}$ & $\tilde{\Phi}_{4}$ & $\Phi_5$ & $\Phi_6$ & $\hat{\Phi}_{3}$ & $\hat{\Phi}_{5}$ \\
\midrule
$\surd$ & 0 & 0 & 0 & 0 & 1 & 0 & 0 & 0 & 0 & 0 & 0 & 1 & 0 & 0 & 0 & 0 \\
 & 0 & 0 & 0 & 0 & 1 & 0 & 0 & 0 & 0 & 0 & 1 & 0 & 0 & 0 & 0 & 0 \\
$\surd$ & 0 & 0 & 0 & 1 & 0 & 0 & 0 & 0 & 0 & 0 & 0 & 0 & 0 & 0 & 1 & 1 \\
$\surd$ & 0 & 0 & 0 & 1 & 0 & 0 & 0 & 0 & 0 & 1 & 0 & 0 & 1 & 0 & 0 & 0 \\
$\surd$ & 0 & 0 & 1 & 0 & 0 & 0 & 0 & 0 & 1 & 0 & 0 & 1 & 0 & 0 & 0 & 0 \\
 & 0 & 0 & 1 & 0 & 0 & 0 & 0 & 0 & 1 & 0 & 1 & 0 & 0 & 0 & 0 & 0 \\
 & 0 & 0 & 1 & 0 & 0 & 0 & 0 & 1 & 0 & 0 & 0 & 1 & 0 & 0 & 0 & 0 \\
$\surd$ & 0 & 0 & 1 & 0 & 0 & 0 & 0 & 1 & 0 & 0 & 1 & 0 & 0 & 0 & 0 & 0 \\
$\surd$ & 0 & 1 & 0 & 0 & 0 & 0 & 0 & 0 & 0 & 0 & 0 & 0 & 0 & 1 & 1 & 0 \\
$\surd$ & 0 & 1 & 0 & 0 & 0 & 0 & 1 & 0 & 0 & 1 & 0 & 0 & 0 & 0 & 0 & 0 \\
 & 1 & 0 & 0 & 0 & 0 & 1 & 0 & 0 & 1 & 0 & 0 & 0 & 0 & 0 & 0 & 0 \\
$\surd$ & 1 & 0 & 0 & 0 & 0 & 1 & 0 & 1 & 0 & 0 & 0 & 0 & 0 & 0 & 0 & 0 \\
\bottomrule
\end{tabular}
\caption{Half-BPS monopole operators in the 3d theory for $(A_{1},A_{5})$. Each row indicates the dressed monopole operator with magnetic charge $m$, dressed by each chiral by the power in the corresponding column. Operators included in the superpotential are indicated with a check mark.}
\label{tab:a5mono}
\end{table}

There are 12 dressed monopoles with $R=2$ that are neutral under $U(1)_{f}$, listed in Table~\ref{tab:a5mono}.
The global symmetry before turning on the superpotential is $U(1)^{10}$, and we expect that there exists a superpotential deformation that
breaks all the global symmetries except for the $U(1)_f$ and the $U(1)_A$ symmetry in the IR. Among all possible deformations, we find that turning on the monopole operators indicated by checkmark in Table \ref{tab:a5mono} preserves the following symmetry:
\begin{align}
	U(1)_{f}=U(1)_{T_{1}}-U(1)_{T_{6}},\qquad U(1)_{A}=\frac12\left(U(1)_{T_{1}}+U(1)_{T_{6}}\right)-U(1)_{\Phi_{2}}+U(1)_{\Phi_{4}}
\end{align}
Using the gauge symmetry of the fourth node, the axial symmetry can also be represented as 
\begin{align}
U(1)_A = \frac12\left(U(1)_{T_{1}}+U(1)_{T_{6}}\right)-U(1)_{\Phi_{2}}-U(1)_{\widetilde\Phi_{4}} + U(1)_{T_3} + U(1)_{T_5} \ .
\end{align}

The $F$-maximization is not feasible for this example, but one can start from the assumption that the Schur index \eqref{schur A5}, when interpreted as the half-index of the 3d $\CN=4$ theory, is computed in the A-twisted limit and reverse engineer it to find the superconformal R-charge. The superconformal index computed with these data is
\begin{align}
\begin{split}
	\mathcal{I}_{SCI}(\mathfrak{q},T,z)&=1-T\mathfrak{q}^{\frac{1}{2}}+T^{\frac{3}{2}}(z^{-1}+z)\mathfrak{q}^{\frac{3}{4}}+(-2+T^2)\mathfrak{q}\\
	&\qquad +(-T^{\frac{1}{2}}+T^{\frac{5}{2}})(z^{-1}+z)\mathfrak{q}^{\frac{5}{4}}+(T^{-1}+T^3(z^{-1}+1+z^2))\mathfrak{q}^{\frac{3}{2}}+\cdots ,\\
    \mathcal{I}_{C}(\mathfrak{q},z)&=1,\\
	\mathcal{I}_{H}(T,z)&=1+T+(z^{-1}+z)T^{\frac{3}{2}}+T^2+(z^{-1}+z)T^{\frac{5}{2}}+(z^{-2}+1+z^2)T^3+\cdots .
\end{split}
\end{align}
The superconformal index contains the term $T\mathfrak{q}^{\frac{1}{2}}$, which can be thought of as the contribution from the $U(1)_{f}$ moment map operator.
The Higgs limit coincides with the Hilbert series of $\mathbb{C}^2 /\mathbb{Z} _{3}$ given as
\begin{align}
	\mathrm{HS}(\mathbb{C} ^2 /\mathbb{Z} _{3};z,T)=\frac{1-T^3}{(1-T)(1-z T^{\frac{3}{2}})(1-z^{-1}T^{\frac{3}{2}})} \ . 
\end{align} 
In fact, among all possible deformations, we find that our choice presented above is the only one that reproduces the correct Higgs branch Hilbert series. 

Now, we can calculate the $T$-refined half-index in the A-twisted limit and find that it matches the proposed Macdonald index of the $(A_{1},A_{5})$ AD theory 
\begin{align} \label{eq:a5Mac}
\begin{split}
	\II &=\frac{1}{(q)_{\infty}^{6}}\sum _{m\in\mathbb{Z} ^{6}}q^{\frac{1}{2}m^T K m}(q^{\frac{1}{2}})^{m_{1}+m_{6}+2(m_{2}+m_{3}+m_{4}+m_{5})}z^{m_{1}-m_{6}} T^{\frac{1}{2}(m_{1}+m_{6})+m_{3}+m_{5}} \\
			   &\quad \times \prod_{i=1}^6(q^{1+m_{i}})_\infty \cdot (q^{1+m_{2}}T)_\infty(q^{1+m_{4}}T)_\infty (q^{1+m_{1}+m_{3}-m_{6}})_\infty(q^{1-m_{1}+m_{5}+m_{6}})_\infty\\
	&=(q)_\infty^{2}(qT)_\infty^2 \sum_{m\in \mathbb{Z} ^{6}} \frac{q^{\frac{1}{2} m^T K m}(q^{\frac{1}{2}})^{m_{1}+m_{6}+2(m_{2}+m_{3}+m_{4}+m_{5})}z^{m_{1}-m_{6}}T^{\frac{1}{2}(m_{1}+m_{6})+m_{3}+m_{5}}}
	{\prod _{i=1}^6(q)_{m_{i}} \cdot (qT)_{m_{2}}(qT)_{m_{4}}(q)_{m_{1}+m_{3}-m_{6}}(q)_{-m_{1}+m_{5}+m_{6}} }
\end{split}
\end{align}
Expanding in $q$-series, we obtain
\begin{align}
\begin{split}
	\II &=1+q T+q^{3/2} T^{3/2} \left(z+\frac{1}{z}\right)+q^2 \left(T^2+2 T\right) \\
    &\quad +q^{5/2} \left(T^{5/2} \left(z+\frac{1}{z}\right)+T^{3/2} \left(z+\frac{1}{z}\right)\right) +q^3 \left(T^3 \left(z^2+\frac{1}{z^2}+1\right)+2 T^2+2 T\right)\\
     &\quad +q^{7/2} \left(T^{7/2} \left(z+\frac{1}{z}\right)+T^{5/2} \left(2 z+\frac{2}{z}\right)+T^{3/2} \left(z+\frac{1}{z}\right)\right)\\
	 &\quad +q^4 \left(T^4 \left(z^2+\frac{1}{z^2}+1\right)+T^3 \left(z^2+\frac{1}{z^2}+2\right)+5 T^2+2 T\right)+\cdots, 
\end{split}
\end{align}
which agrees with the known result. 

\subsubsection{$A_{7}$}
The formula \eqref{eq:a2n1Schur} gives
\begin{align}
I_{\text{Schur}}^{A_{7}}=(q)_\infty^{6}\sum _{m\in\mathbb{Z} ^{8}}\frac{q^{\frac{1}{2}\sum _{i=1}^{7}b^{A_{7}m_{i}m_{j}}}q^{\sum _{i=1}^{7}m_{i}}z^{m_{1}-m_{8}}}{\prod _{i=1}^{8}(q)_{m_{i}}\prod _{j=1}^{3}(q)_{m_{2j}}\prod_{k=1}^{3}(q)_{(-1)^{k+1}m_{1}+m_{2k+1}+(-1)^{k}m_{8}}}.
\end{align} 
From this, we can read off the corresponding 3d theory as before: a $U(1)^8$ ACSM theory with 14 chiral multiplets. 
This gauge theory has a rather high rank, which makes the direct computation of the superconformal index infeasible. Instead, as a non-trivial consistency check, we show that there exists a suitable superpotential deformation and a corresponding $U(1)_A$ symmetry that reproduces the conjecutural form of the Macdonald index.

We conjecture that there exists the following superpotential deformation
\begin{align}
\begin{split}
    \mathcal{W}&=
    V_{1,8} \tilde{\Phi }_2
    +V_2 \Phi _1 \Phi _3
    +V_2 \Phi _8 \hat{\Phi }_3
    +V_3 \Phi _2 \Phi _4
    +V_3 \tilde{\Phi }_2 \tilde{\Phi }_4
    +V_4 \Phi _5 \Phi _3\\
    &\qquad+V_4 \hat{\Phi }_3 \hat{\Phi }_5
    +V_5 \Phi _4 \Phi _6
    +V_5 \tilde{\Phi }_4 \tilde{\Phi }_6
    +V_6 \Phi _5 \Phi _7
    +V_6 \hat{\Phi }_5 \hat{\Phi }_7
    +V_7 \tilde{\Phi }_6
\end{split}
\end{align}
and the theory flows to an $\CN=4$ SCFT in the IR with a trivial Coulomb branch and the Higgs branch given as $\mathbb{C}^2/\mathbb{Z}_4$. 
Here $V_{i}$ denotes a bare monopole charged $+1$ under $U(1)_{T_{i}}$, and $V_{i,j}$ denotes the one charged under both $U(1)_{T_{i}}$ and $U(1)_{T_j}$ by $+1$.
We justify this by reproducing the Macdonald index from the half-index by identifying the flavor and $U(1)_A$ symmetry as
\begin{align}
	U(1)_{f}=U(1)_{T_{1}}-U(1)_{T_{8}},\quad U(1)_{A}= -U(1)_{\tilde{\Phi}_2} -U(1)_{{\Phi}_4} -U(1)_{\tilde{\Phi}_6} + \sum_{i=1}^4U(1)_{T_{2i-1}} \ . 
\end{align}

Computing the $T$-refined half-index in the A-twisted limit, we get
\begin{align} \label{eq:a7Mac}
\begin{split}
	\II(q,T,z) &=\frac{1}{(q)^{8}_{\infty }}\sum _{m\in\mathbb{Z} ^{8}}q^{\frac{1}{2} m^{t}Km}q^{\sum _{i=1}^{7}m_{i}}z^{m_{1}-m_{8}}T^{m_{1}+m_{3}+m_{5}+m_{7}}\\
	&\quad \times \prod _{i=1}^8(q^{1+m_{i}})_\infty\prod _{j=1}^{3}(q^{1+m_{2j}}T)_\infty\prod _{k=1}^{3}(q^{1+(-1)^{k}m_{1}-m_{2k+1}+(-1)^{k+1}m_{8}})_\infty \\
&=(q)_\infty^{3}(qT)_\infty^3\sum _{m\in\mathbb{Z} ^{8}}\frac{q^{\frac{1}{2}\sum _{i=1}^{7}b^{A_{7}m_{i}m_{j}}}q^{\sum _{i=1}^{7}m_{i}}z^{m_{1}-m_{8}}T^{m_{1}+m_{3}+m_{5}+m_{7}}}{\prod _{i=1}^{8}(q)_{m_{i}}\prod _{j=1}^{3}(qT)_{m_{2j}}\prod_{k=1}^{3}(q)_{(-1)^{k+1}m_{1}+m_{2k+1}+(-1)^{k}m_{8}}} \ , 
\end{split}
\end{align}
which indeed agrees with our proposed formula. 
Upon expanding the above expression in $q$-series, we obtain
\begin{align}
\begin{split}
\II &=
1 +q T +q^2 \left(T^2 \left(z+\frac{1}{z}+1\right)+2 T\right) \\
&\quad +q^3 \left(T^3 \left(z+\frac{1}{z}+1\right)+T^2 \left(z+\frac{1}{z}+2\right)+2 T\right)\\
&\quad +q^4 \left(T^4 \left(z^2+\frac{1}{z^2}+z+\frac{1}{z}+1\right)+T^3 \left(2 z+\frac{2}{z}+2\right)+T^2 \left(z+\frac{1}{z}+5\right)+2 T\right) + \cdots. 
\end{split}
\end{align}
This agrees with the known expression for the Macdonald index. 

\subsubsection{General Form}
Based on the above results, we conjecture that the following superpotential triggers an RG flow to $\CN=4$ SCFT in the IR:
\begin{align}
\begin{split}
    W &= V_{1,2n+2}\widetilde{\Phi}_{2}+V_{2}\hat{\Phi}_{2n+1}\Phi_{2n+2}
	+\sum _{i=1}^{n}V_{2i}\Phi_{2i-1}\Phi_{2i+1}\\
	&\qquad +\sum _{i=2}^{n}V_{2i}\hat{\Phi}_{2i-1}\hat{\Phi}_{2i+1}  +\sum _{i=2}^{n}V_{2i-1}\Phi_{2i-2}\Phi_{2i}+\sum_{i=2}^{n+1}V_{2i-1}\tilde{\Phi}_{2i-2}\tilde{\Phi}_{2i} \ , 
\end{split}	
\end{align}
where $V_{i}$ denotes a bare monopole charged $+1$ under $U(1)_{T_{i}}$, and $V_{i,j}$ denotes the one charged under both $U(1)_{T_{i}}$ and $U(1)_{T_j}$ by $+1$. Here, we set $\tilde{\Phi}_{2n+2}=1$ for brevity.
The axial symmetry that survives this deformation is
\begin{align}
    U(1)_A =
    \frac{1}{2}(U(1)_{T_1}+U(1)_{T_{2n+2}})+\sum_{i=1}^n U(1)_{T_{2i+1}}-\sum_{j=1}^{n-1}(U(1)_{\Phi_{2i}}-U(1)_{\tilde{\Phi}_{2i+2}}) \quad (\text{$n$ even}) \ , 
\end{align}
for $n$ even, and 
\begin{align}
    U(1)_A = \sum_{i=0}^{n}U(1)_{T_{2i+1}}-\sum_{j=1}^{n-2}(U(1)_{\tilde\Phi_{2i}}-U(1)_{\Phi_{2i+2}})-U(1)_{\tilde\Phi_{2n}} \quad (\text{$n$ odd}) \ , 
\end{align}
for $n$ odd. 
The refined half-index with this $U(1)_A$ symmetry leads to the following form of the Macdonald index, which reproduces the conjecture in section \ref{sec: conjectural Macdonald}:
\begin{align}
	I_{\text{Mac}}^{A_{2n+1}}=(q)_{\infty }^{n}(qT)_\infty^n\sum _{k,\ell \in \mathbb{Z} ^{2n+1}}& \frac{q^{\frac{1}{2}k^{t}b^{A_{2n+1}}\ell}q^{\frac12\sum _{i}^{2n+1}(k_{i}+\ell_{i})}z^{\ell_{1}-k_{1}} T^{\frac{1}{2} \sum_{i=1}^{n+1}(k_{2i-1}+\ell_{2i-1})}}{\prod_{i=1}^{n+1}(q)_{k_{2i-1}}(q)_{\ell_{2i-1}}\prod _{j=1}^{n}(qT)_{k_{2j}}(q)_{\ell_{2j}}} \nonumber \\
	&\times \prod_{i=1}^{n}\delta_{(-1)^{i+1}k_{1}+k_{2i+1},(-1)^{i+1}\ell_{1}+\ell_{2i+1}}\prod_{j=1}^{n}\delta_{k_{2j},\ell_{2j}} \ . 
\end{align}

\subsection{$(A_1, D_{2n+1})$ theory}
\begin{figure}
  \centering
  \begin{tikzpicture}
			\node[W] (1) at (0,0){};
		 	\node[above=3mm] at (1) {$\gamma_1$};
            
            \node[W] (2) at (2,0){};
		 	\node[above=3mm] at (2) {$\gamma_2$};

            \node[] (3) at (4,0){$\cdots$};
            
            \node[W] (4) at (6,0){};
		 	\node[above=3mm] at (4) {$\gamma_{2n-1}$};

            \node[W] (5) at ({6+\dist*cos(30)},{0+\dist*sin(30)}){};
		 	\node[above=3mm] at (5) {$\gamma_{2n}$};

            \node[W] (6) at ({6+\dist*cos(30)},{0-\dist*sin(30)}){};
		 	\node[above=3mm] at (6) {$\gamma_{2n+1}$};
            
			\draw[->] (1)--(2);
            \draw[<-] (2)--(3);
            \draw[<-] (3)--(4);
            \draw[->] (4)--(5);
            \draw[->] (4)--(6);
  \end{tikzpicture}
  \caption{\label{fig:d2n1quiver}
  BPS quiver for the $(A_1, D_{2n+1})$ theory in the canonical chamber}
\end{figure}
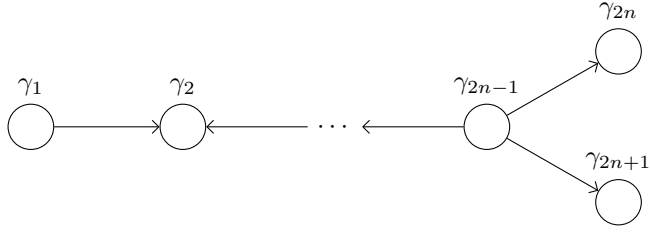

The $(A_1, D_{2n+1})$ theory has an $n$-dimensional Coulomb branch and a Higgs branch given by $\mathbb{C}^2/\mathbb{Z}_2$. It has $SU(2)$ flavor symmetry, and the associated VOA is given by $\hat{\mathfrak{su}(2)}_{-\frac{4n}{2n+1}}$. 
The BPS quiver for this theory is given in Figure~\ref{fig:d2n1quiver}. 
The $SU(2)$ flavor symmetry corresponds to
\begin{align}
	\gamma_{f}=\gamma_{2n+1}-\gamma_{2n},\qquad \mathrm{Tr} [X_{\gamma_{f}}]=z^2.
\end{align} 
The trace formula \eqref{eq:traceSchur} gives
\begin{align}\label{eq:d2n1Schur}
\begin{split}
	I_{\text{Schur}}^{D_{2n+1}}(q,T,z)=
	(q)_{\infty }^{2n}
	\sum _{k,\ell \in \mathbb{Z} ^{2n+1}}& 
	\frac{
	q^{\frac{1}{2}k^{t}b^{D_{2n+1}}\ell}
	q^{\frac12\sum _{i}^{2n+1}(k_{i}+\ell_{i})}z^{\ell_{2n+1}-k_{2n+1}}}
	{(q)_{k_{2n+1}}(q)_{\ell_{2n+1}}\prod_{i=1}^{2n}(qT)_{k_{i}}(q)_{\ell_{i}}}\\
	&\times \left(\prod_{i=1}^{2n-1}\delta_{k_{i},\ell_{i}}\right)\delta_{k_{2n}+k_{2n+1},\ell_{2n}+\ell_{2n+1}}\ .
\end{split}
\end{align} 
From this, we extract $U(1)^{2n+2}$ Chern-Simons matter theory with level
\begin{align}
    K=\left( 
	\begin{array}{ccc|c}
		\multicolumn{3}{c}{\multirow{3}{*}{\raisebox{-1mm}{\scalebox{1.1}{$b^{D_{2n+1}}$}}}}\vrule &0\\
		& & &\vdots\\
		& & & 0\\
		\hline
		0& \cdots  & 0 & 0
	\end{array}
	\right) \ . 
\end{align}
The mixed Chern-Simons levels between the gauge $U(1)$'s and $R$- symmetry are $\rho=\{2,2,\cdots,2,0\}$. 
The Cartan of the flavor symmetry is identified with
\begin{align}
\begin{split}
    U(1)_f=U(1)_{T_{2n+2}}-U(1)_{T_{2n}}.
\end{split}
\end{align}
We have $4n+2$ chiral multiplets, all with $R$-charge $0$.
Their gauge charges are given as follows:
\begin{itemize}
\item $\Phi_i$ has charge $-1$ under $U(1)_i$ for $i=1,2,3,\cdots,2n+2$.
\item $\tilde{\Phi}_j$ has charge $-1$ under $U(1)_j$ for $j=1,2,3,\cdots,2n-1$.
\item $X$ has charge $-1$ under $U(1)_{2n}$, $-1$ under $U(1)_{2n+1}$ and $+1$ under $U(1)_{2n+2}$.
\end{itemize}

\subsubsection{$D_{3}$}
For $n=1$, the formula \eqref {eq:d2n1Schur} gives
\begin{align}
	I_{\text{Schur}}^{D_{3}}(q,z)=
	(q)_{\infty }^{2}\sum _{m\in \mathbb{Z} ^{4}}\frac{q^{m_{1}(m_{2}+m_{3})}q^{m_{1}+m_{2}+m_{3}}z^{m_{4}-m_{2}}}{(q)_{m_{1}}(q)_{m_{1}}(q)_{m_{2}}(q)_{m_{3}}(q)_{m_{4}}(q)_{m_{2}+m_{3}-m_{4}}}
\end{align} 
from which we can read off the ACSM theory with $U(1)^4$ gauge group and 6 chiral multiplets. 
\begin{table}[htb]
\centering
\begin{tabular}{@{}c cccc| ccccccc@{}}
\toprule
 & $m_1$ & $m_2$ & $m_3$ & $m_4$ & $\Phi_1$ & $\tilde{\Phi}_1$ & $\Phi_2$ & $\Phi_3$ & $\Phi_4$ & $X$ \\
\midrule
 & 0 & 0 & 1 & 0 & 0 & 1 & 0 & 0 & 0 & 0 \\
$\surd$ & 0 & 0 & 1 & 0 & 1 & 0 & 0 & 0 & 0 & 0 \\
 & 0 & 1 & 0 & 1 & 0 & 1 & 0 & 0 & 0 & 0 \\
$\surd$ & 0 & 1 & 0 & 1 & 1 & 0 & 0 & 0 & 0 & 0 \\
$\surd$ & 1 & 0 & 0 & 0 & 0 & 0 & 0 & 0 & 1 & 1 \\
$\surd$ & 1 & 0 & 0 & 0 & 0 & 0 & 1 & 1 & 0 & 0 \\
\bottomrule
\end{tabular}
\caption{Half-BPS dressed monopole operators for the 3d theory of $(A_{1},D_{3})$. Operators included in the superpotential are indicated with a check mark.}
\label{tab:d3mono}
\end{table}

There are $6$ monopoles with $R=2$ and preserving $U(1)_{f}$, listed in Table~\ref{tab:d3mono}.
The global symmetry before turning on the superpotential is $U(1)^{6}$. We turn on the superpotential given in Table \ref{tab:d3mono} to break the global symmetry down to $U(1)_f \times U(1)_A$.

Then, we can single out the combinations (up to symmetry $U(1)_{2}\leftrightarrow U(1)_{3}$ and $\Phi_{1}\leftrightarrow \tilde{\Phi}_{1}$)
\begin{align}
	U(1)_{f}=U(1)_{T_{4}}-U(1)_{T_{2}},\quad U(1)_{A}=U(1)_{\tilde{\Phi}_{1}}
\end{align} 
with the superpotential consisting of monopole operators in Table~\ref{tab:d3mono}.
Upon mixing with gauge symmetry, the axial charge can also be represented as
\begin{align}\label{eq:d3axial}
    U(1)_A=U(1)_{T_2}+U(1)_{T_3}-U(1)_{\Phi_1}
\end{align}

The superconformal index and its Coulomb/Higgs limits with this choice are
\begin{align}
\begin{split}
	\mathcal{I}_{SCI}(\mathfrak{q},T,z) &=1+\chi^{SU(2)}_{\mathbf{3}}(z)T\mathfrak{q}^{\frac{1}{2}}+\left(-\chi^{SU(2)}_{\mathbf{3}}(z)-1+T^2\chi^{SU(2)}_{\mathbf{5}}(z)\right)\mathfrak{q}\\
	&+\left( T+\frac{1}{T}+T^{3}\chi^{SU(2)}_{\mathbf{7}}(z)-T\chi^{SU(2)}_{\mathbf{5}}(z) \right)\mathfrak{q}^{\frac{3}{2}}+\cdots,\\
    \mathcal{I}_{C}(\mathfrak{q},z)&=1,\\
	\mathcal{I}_{H}(T,z)&=1+\chi_{\mathbf{3}}^{SU(2)}T+\chi_{\mathbf{5}}^{SU(2)}T^2+\chi_{\mathbf{7}}^{SU(2)}T^3+\cdots .
\end{split}
\end{align}
In the superconformal index, $\chi^{SU(2)}_{\mathbf{3}}\mathfrak{q}^{\frac{1}{2}}$ comes from the $SU(2)_{f}$ moment map and the extra  $-\mathfrak{q}$ corresponds  to the $\mathcal{N}=4$ stress-energy tensor.
The Coulomb and Higgs limits indeed reproduce the Hilbert series of the trivial Coulomb branch and Higgs branch of $\mathbb{C} ^2/\mathbb{Z} _{2}$, which is given as
\begin{align}
	\mathrm{HS}(\mathbb{C} ^2/\mathbb{Z} _{2};T,z)=\sum _{i=0}^{\infty }\chi_{\mathbf{2i+1}}^{SU(2)}(z)T^{i} \ . 
\end{align} 

\begin{figure}[htpb]
	\centering
	\includegraphics[width=0.65\textwidth]{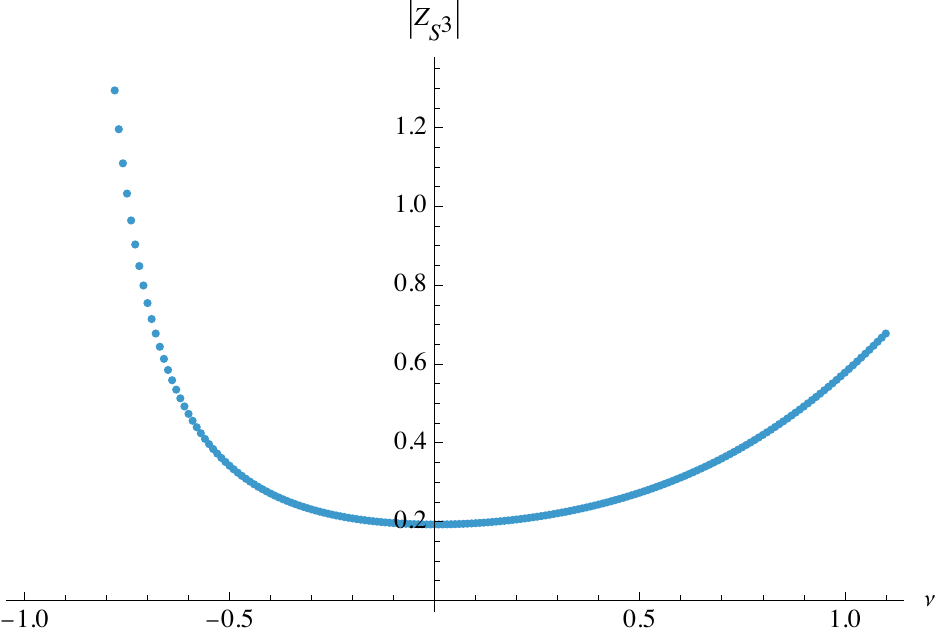}
\caption{The absolute value of three-sphere partition function for the 3d theory for $(A_{1},D_{3})$ as a function of the mixing parameter $\nu$ with $U(1)_{A}=U(1)_{\tilde{\Phi}_{1}}$. The function $|Z_{S^{3}}|$ is minimized at $\nu=0$}
	\label{fig:d3Fmax}
\end{figure}

We also evaluate the three-sphere partition function as a function of the mixing parameter $\nu$ and indeed find that it is minimized at $\nu=0$, see Figure~\ref{fig:d3Fmax}. This shows that our identification of superconformal $R$-symmetry and $U(1)_A$ symmetry is indeed correct.

The $T$-refined half-index in the A-twisted limit with \eqref{eq:d3axial} is given as
\begin{align} \label{eq:d3Mac}
\begin{split}
	\II(q,T,z) &=\frac{1}{(q)_\infty^{4}}\sum _{m\in\mathbb{Z}^{4}}q^{m_{1}(m_{2}+m_{3})}q^{m_{1}+m_{2}+m_{3}}z^{m_{4}-m_{2}}T^{m_{2}+m_{3}} \\
	&\qquad \times \left(\prod _{i=1}^{4}(q^{1+m_{i}})_\infty\right) (q^{1+m_{1}}T)_\infty(q^{1+m_{2}+m_{3}-m_{4}})_\infty \\
	&=(q)_\infty(qT)_\infty\sum _{m\in \mathbb{Z} ^{4}}\frac{q^{m_{1}(m_{2}+m_{3})}q^{m_{1}+m_{2}+m_{3}}z^{m_{4}-m_{2}}T^{m_{2}+m_{3}}}{\left( \prod _{i=1}^{4}(q)_{m_{i}} \right)(qT)_{m_{1}}(q)_{m_{2}+m_{3}-m_{4}}} \ , 
\end{split}
\end{align}
which matches with the proposed formula for the Macdonald index of $(A_{1},D_{3})$ AD theory. Upon series expansion, we obtain
\begin{align}
\begin{split}
	\II(q,T,z) &=1+qT\chi^{SU(2)}_{\mathbf{3}}(z)+q^2\left( T\chi^{SU(2)}_{\mathbf{3}}(z)+T^2\chi^{SU(2)}_{\mathbf{5}}(z) \right) \\
	 &\qquad +q^3\bigg( T(1+\chi^{SU(2)}_{\mathbf{3}}(z))+T^2(1+2\chi^{SU(2)}_{\mathbf{3}}(z)+2\chi^{SU(2)}_{\mathbf{5}}(z))\\
	 &\qquad+T^3(\chi^{SU(2)}_{\mathbf{5}}(z)+\chi^{SU(2)}_{\mathbf{7}}(z))+T^{4}\chi^{SU(2)}_{\mathbf{9}}(z) \bigg)+\cdots .
\end{split}
\end{align}
This agrees with the previous results. 

\subsubsection{$D_{5}$}
The IR formula for the Schur index \eqref{eq:d2n1Schur} gives
\begin{align}
	I_{\text{Schur}}^{D_{5}}=(q)_\infty^{4}\sum _{m\in\mathbb{Z} ^{6}}
	\frac{q^{m_{1}m_{2}+m_{3}(m_{2}+m_{4}+m_{5})}q^{\sum _{i=1}^{6}m_{i}}z^{m_{6}-m_{4}}}{\prod _{i=1}^{3}(q)_{m_{i}}^2\prod _{i=4}^{6}(q)_{m_{i}}(q)_{m_{4}+m_{5}-m_{6}}} \ . 
\end{align} 
from which we can read off the ACSM theory with $U(1)^6$ gauge group and 10 chiral multiplets. 
We conjecture that there exists a superpotential deformation with 8 monopole operators
\begin{align}
\begin{split}
	\mathcal{W} &=
	V_{1}\tilde{\Phi}_{2} +V_{3}\Phi_{2}\Phi_{6}X+V_{3}\tilde{\Phi}_{2}\Phi_{6}X+V_{3}\tilde{\Phi}_{2}\Phi_{4}\Phi_{5}+V_{5}\tilde{\Phi}_{3} \\
  &\qquad +V_{2}\Phi_{1}\Phi_{3}+V_{2}\tilde{\Phi}_{1}\tilde{\Phi}_{3}+V_{4,6}\tilde{\Phi}_{3}
\end{split}
\end{align} 
leaving the following symmetry unbroken
\begin{align}
    U(1)_A=U(1)_{T_2}+U(1)_{T_4}+U(1)_{T_5}-U(1)_{\Phi_1}-U(1)_{\tilde{\Phi}_3}\ ,
\end{align}
together with $U(1)_f=U(1)_{T_6}-U(1)_{T_4}$.
As before, this proporsal is tested against the $T$-refined superconformal index and its Higgs/Coulomb limits, given as
\begin{align}
\begin{split}
	\mathcal{I}_{\mathrm{SCI}}(\mathfrak{q},T,z)
	&= 1-\chi_{\mathbf{3}}^{SU(2)}(z)\,T\,\mathfrak{q}^{1/2}
	+\Big(-1-\chi_{\mathbf{3}}^{SU(2)}(z)+T^2\chi_{\mathbf{5}}^{SU(2)}(z)\Big)\mathfrak{q}\\
    &\qquad-\Big(T^{-1}+T+T^{3}\chi_{\mathbf{7}}^{SU(2)}(z)-T\big(\chi_{\mathbf{3}}^{SU(2)}(z)+\chi_{\mathbf{5}}^{SU(2)}(z)\big)\Big)\mathfrak{q}^{3/2}+\cdots,\\
	\mathcal{I}_{C}(\mathfrak{q},z)&=1,\\
    \mathcal{I}_{H}(T,z)
    &=1+\chi_{\mathbf{3}}^{SU(2)}(z)\,T+\chi_{\mathbf{5}}^{SU(2)}(z)\,T^{2}+\chi_{\mathbf{7}}^{SU(2)}(z)\,T^{3}+\cdots .
\end{split}
\end{align}
The term $\chi^{SU(2)}_{\mathbf{3}}\mathfrak{q}^{\frac{1}{2}}$ in the superconformal index comes from the $SU(2)_{f}$ moment map and the term $-\mathfrak{q}$ corresponds to the $\mathcal{N}=4$ stress-energy tensor.
The Coulomb and Higgs limits correctly reproduce the Hilbert series of the trivial Coulomb branch and $\mathbb{C} ^2/\mathbb{Z}_{2}$ Higgs branch. 

The $T$-refined half-index matches our proposed expression for the Macdonald index of the $(A_{1},D_{5})$ AD theory, given as
\begin{align} \label{eq:d5Mac}
\begin{split}
	\II 
	&=\frac{1}{(q)_\infty^{6}}\sum _{m\in \mathbb{Z} ^{6}}q^{m_{1}m_{2}+m_{2}(m_{2}+m_{4}+m_{5})}q^{\sum_{i=1}^{6} m_{i}}z^{m_{6}-m_{4}}T^{m_{2}+m_{4}+m_{5}}\\
    &\qquad \times(q^{1+m_{1}}T)_\infty(q^{1+m_{1}})_\infty(q^{1+m_{2}})_\infty^2(q^{1+m_{3}}T)_\infty(q^{1+m_{3}})_\infty(q^{1+m_{4}})_\infty \\
    &\qquad \times (q^{1+m_{5}})_\infty (q^{1+m_{6}})_\infty(q^{1+m_{4}+m_{5}-m_{6}})_\infty \\ 
	&= (q)_{\infty }^{2}(qT)_\infty^2\sum _{m\in \mathbb{Z} ^{6}}\frac{q^{m_{1}m_{2}+m_{2}(m_{2}+m_{4}+m_{5})}q^{\sum_{i=1}^{6} m_{i}}z^{m_{6}-m_{4}}T^{m_{2}+m_{4}+m_{5}}}{(qT)_{m_{1}}(q)_{m_{1}}(q)_{m_{2}}^{2}(qT)_{m_{3}}(q)_{m_{3}}(q)_{m_{4}}(q)_{m_{5}}(q)_{m_{6}}(q)_{m_{4}+m_{5}-m_{6}}} \ .
\end{split}
\end{align}
Upon series expansion in $q$, we obtain
\begin{align}
\begin{split}
	\II(q,T,z)&=1+T\chi^{SU(2)}_{\mathbf{3}}(z)q+\left( T+T\chi^{SU(2)}_{\mathbf{3}}+T^2\chi^{SU(2)}_{\mathbf{5}} \right)q^2+\\
	 &\qquad +\left( T(1+\chi^{SU(2)}_{\mathbf{3}}(z))+T^2(\chi^{SU(2)}_{\mathbf{5}}(z)+2\chi^{SU(2)}_{\mathbf{3}}(z))+T^3\chi^{SU(2)}_{\mathbf{7}}(z) \right)q^3+\cdots ,
\end{split}
\end{align}
which agrees with known results. 

\subsubsection{General Form}
Based on the expressions \eqref{eq:d3Mac} and \eqref {eq:d5Mac}, we turn on the superpotential for the $U(1)^{2n+2}$ ACSM theory as follows:
\begin{align}
\begin{split}
W&=
	V_{1}\Phi_{2}+\sum _{i=2}^{2n-2}V_{i}\Phi_{i-1}\Phi_{i+1}+\sum_{i=2}^{2n-2}V_{i}\tilde{\Phi}_{i-1}\tilde{\Phi}_{i+1}\\
	&\qquad +V_{2n-1}\Phi_{2n-2}\Phi_{2n+2}X+V_{2n-1}\tilde{\Phi}_{2n-2}\Phi_{2n+2}X+V_{2n-1}\tilde{\Phi}_{2n-2}\Phi_{2n}\Phi_{2n+1}\\
	&\qquad +V_{2n,2n+2}\tilde{\Phi}_{2n-1}+V_{2n+1}\tilde{\Phi}_{2n-1}
\end{split}
\end{align}
We conjecture that this theory flows to an $\mathcal{N}=4$ SCFT in the IR also with enhanced $SU(2)$ flavor symmetry. In particular, we find that axial $U(1)_A \subset SO(4)_R$ symmetry comes from the following combinations of UV symmetries:
\begin{align}
   U(1)_A=
\begin{cases}
    \displaystyle \sum_{i=1}^nU(1)_{T_{2i}}+U(1)_{T_{2n+1}}-\sum_{i=1}^{n-2}(U(1)_{\tilde{\Phi}_{2i-1}}+U(1)_{\Phi_{2i+1}})-U(1)_{\tilde{\Phi}_{2n-1}} & \text{($n$ odd)}\\
    \displaystyle \sum_{i=1}^n U(1)_{T_{2i}}+U(1)_{T_{2n+1}}-\sum_{i=1}^{n-1}(U(1)_{\Phi_{2i-1}}+U(1)_{\tilde\Phi_{2i+1}}) & \text{($n$ even)}
\end{cases}
\end{align}

With these at hand, we compute the $U(1)_A$ refined half-index of this theory to obtain
\begin{align}
\begin{split}
	&I_{\text{Mac}}^{D_{2n+1}}(q,T,z)=
	(q)_{\infty }^{n}(qT)_\infty^n\\
	&\qquad\times\sum _{k,\ell \in \mathbb{Z} ^{2n+1}}
	\frac{
	q^{\frac{1}{2}k^{t}b^{D_{2n+1}}\ell}
	q^{\frac12\sum _{i}^{2n+1}(k_{i}+\ell_{i})}z^{\ell_{2n+1}-k_{2n+1}}
	(T^{\frac{1}{2}})^{\sum _{i=1}^{n}(k_{2i}+\ell_{2i})+k_{2n+1}+\ell_{2n+1}}}
	{(q)_{k_{2n+1}}(q)_{\ell_{2n+1}}\prod_{i=1}^{n}(qT)_{k_{2i-1}}(q)_{\ell_{2i-1}}\prod _{j=1}^{n}(q)_{k_{2j}}(q)_{\ell_{2j}}}\\
	&\qquad\qquad\times \left(\prod_{i=1}^{2n-1}\delta_{k_{i},\ell_{i}}\right)\delta_{k_{2n}+k_{2n+1},\ell_{2n}+\ell_{2n+1}} \ , 
\end{split}
\end{align} 
which agrees with our conjectural formula for the Macdonald index of the $(A_{1},D_{2n+1})$ AD theory. Indeed, the Macdonald grading in 4d corresponds to the $U(1)_{A}$ charge of the 3d theory.

\subsection{$(A_1, D_{2n+2})$ theory} \label{subsec:Deven}
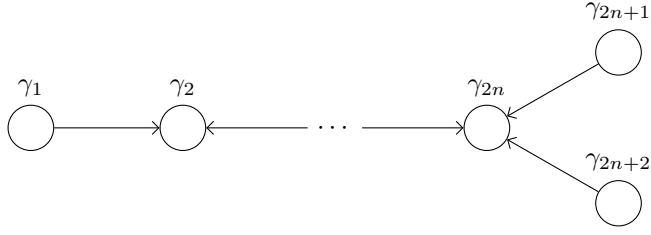
\begin{figure}
  \centering
  \begin{tikzpicture}
			\node[W] (1) at (0,0){};
		 	\node[above=3mm] at (1) {$\gamma_1$};
            
            \node[W] (2) at (2,0){};
		 	\node[above=3mm] at (2) {$\gamma_2$};

            \node[] (3) at (4,0){$\cdots$};
            
            \node[W] (4) at (6,0){};
		 	\node[above=3mm] at (4) {$\gamma_{2n}$};

            \node[W] (5) at ({6+\dist*cos(30)},{0+\dist*sin(30)}){};
		 	\node[above=3mm] at (5) {$\gamma_{2n+1}$};

            \node[W] (6) at ({6+\dist*cos(30)},{0-\dist*sin(30)}){};
		 	\node[above=3mm] at (6) {$\gamma_{2n+2}$};
            
			\draw[->] (1)--(2);
            \draw[<-] (2)--(3);
            \draw[->] (3)--(4);
            \draw[<-] (4)--(5);
            \draw[<-] (4)--(6);
  \end{tikzpicture}
  \caption{\label{fig:d2n2quiver}
  BPS quiver for the $(A_1, D_{2n+2})$ theory in the canonical chamber}
\end{figure}

$(A_{1},D_{2n+2})$ Argyres-Douglas theory has an $n$-dimensional Coulomb branch, and the BPS quiver is given as in Figure~\ref{fig:d2n2quiver}.
It has an $SU(3)$ flavor symmetry for $n=1$, and $SU(2)\times U(1)$ flavor symmetry for $n>1$.
The lattice vectors for the flavor symmetry are given as
\begin{align}
	\gamma_{f_{1}}=\gamma_{2n+2}-\gamma_{2n+1},\quad \gamma_{f_{2}}=\sum _{i=0}^{n}(-1)^{i}\gamma_{2i+1}.
\end{align} 
The IR trace formula for the Schur index \eqref{eq:traceSchur} gives
\begin{align}\label{eq:d2nSchur}
\begin{split}
	&I_{\text{Schur}}^{D_{2n+2}}(q,T,z_{1},z_{2})\\
	&\quad = (q)_{\infty }^{2n}
	\sum _{k,\ell \in \mathbb{Z} ^{2n+2}} 
	\frac{
	q^{\frac{1}{2}k^{t}b^{D_{2n+2}}\ell}
	q^{\frac12\sum _{i}^{2n+2}(k_{i}+\ell_{i})}
	(\mathrm{Tr} [X_{\gamma_{f_{1}}}])^{\ell_{2n+2}-k_{2n+2}}(\mathrm{Tr} [X_{\gamma_{f_{2}}}])^{\ell_{1}-k_{1}}
	}
	{(q)_{k_{2n+1}}(q)_{k_{2n+2}}(q)_{\ell_{2n+1}}(q)_{\ell_{2n+2}}\prod_{i=1}^{2n}(q)_{k_{i}}(q)_{\ell_{i}}}\\
	&\qquad \times
	\left(\prod_{i=1}^{n}\delta_{k_{2i},\ell_{2i}}\right)
	\left(\prod _{j=1}^{n-1}\delta_{(-1)^{j+1}k_{1}+k_{2j+1},(-1)^{j+1}\ell_{1}+\ell_{2j+1}}\right)\\
	&\qquad \times \delta_{(-1)^{n+1}k_{1}+k_{2n+1}+k_{2n+2},(-1)^{n+1}\times \ell_{1}+\ell_{2n+1}+\ell_{2n+2}}.
\end{split}	
\end{align} 
with 
\begin{align}
	\mathrm{Tr} [X_{\gamma_{f_{1}}}]=w^2,\qquad \mathrm{Tr} [X_{\gamma_{f_{2}}}]=
	\begin{cases}
		wx & n \text{ odd}\\ x/w & n \text{ even}
	\end{cases}
\end{align} 
and specifically for $D_{4}$ we set 
\begin{align}
	w^2=z_{1}^2z_{2}^{-1},\qquad w x=z_{1}z_{2}^{-2}.
\end{align} 
to write the index in $SU(3)$ fugacities. 

Again, we start from an observation that this formula is identical to the half-index of the 3d $U(1)^{2n+4}$ Chern-Simons matter theory with level
\begin{align}
    K=
		\left( 
	\begin{array}{ccc|cc}
		\multicolumn{3}{c}{\multirow{3}{*}{\raisebox{-1mm}{\scalebox{1.1}{$b^{D_{2n+2}}$}}}}\vrule &0&0\\
		& & &\vdots&\vdots\\
		& & & 0&0\\
		\hline
		0& \cdots  & 0 & 0&0\\
		0& \cdots  & 0 & 0&0\
	\end{array}
	\right) \ . 
\end{align}
The mixed Chern-Simons levels between the gauge $U(1)$s and $R$-symmetry are $\rho=\{2,2,\cdots,2,0,0\}$ for odd $n$ and $\rho=\{1,2,2,\cdots,2,1,0\}$ for even $n$. 

The theory has two flavor symmetries
\begin{align}
\begin{split}
    U(1)_{f_1}&=U(1)_{T_{1}}-U(1)_{T_{2n+3}},\\
    U(1)_{f_2}&=(-1)^n\left(U(1)_{T_1}-U(1)_{T_{2n+3}}\right)-U(1)_{T_{2n+1}}+U(1)_{T_{2n+4}}.
\end{split}
\end{align}

Also, we have $4n+4$ chiral multiplets, all with trial $R$-charge $0$.
Their gauge charges are
\begin{itemize}
\item $\Phi_i$ has charge $-1$ under $U(1)_i$ for $i=1,2,3,\cdots,2n+4$.
\item $\tilde{\Phi}_j$ has charge $-1$ under $U(1)_j$ for $j=2,4,6,\cdots,2n$.
\item $\hat{\Phi}_k$ has charge $(-1)^{\frac{i-1}{2}}$ under $U(1)_1$, $-1$ under $U(1)_k$ and $(-1)^{\frac{i+1}{2}}$ under $U(1)_{2n+3}$ for $k=3,5,7,\cdots,2n-1$.
\item $X$ has charge $(-1)^n$ under $U(1)_{1}$, $-1$ under $U(1)_{2n+1}$, $-1$ under $U(1)_{2n+2}$, $(-1)^{n+1}$ under $U(1)_{2n+3}$ and $+1$ under $U(1)_{2n+4}$.
\end{itemize}

\subsubsection{$D_{4}$}
For $n=1$, the formula \eqref {eq:d2nSchur} gives
\begin{align}
	I_{\text{Schur}}^{D_{4}}(q,z_{1},z_{2})=(q)_\infty^2\sum _{m\in\mathbb{Z} ^{6}}\frac{q^{\frac{1}{2}\sum _{i,j=1}^{4}b_{ij}m_{i}m_{j}}q^{\sum _{i=1}^{4}m_{i}}(\frac{z_{1}^2}{z_{2}})^{m_{1}-m_{6}}(\frac{z_{1}}{z_{2}^2})^{-m_{1}-m_{4}+m_{5}+m_{6}}}{\left(\prod _{i=1}^{6}(q)_{m_{i}}\right)(q)_{m_{2}}(q)_{m_{1}+m_{3}+m_{4}-m_{5}-m_{6}}}.
\end{align} 
from which we can read off the ACSM theory with $U(1)^6$ gauge group and 8 chiral multiplets. 

\begin{table}[tb]
\centering
\begin{tabular}{@{}c cccccc| cccccccc@{}}
\toprule
 & $m_1$ & $m_2$ & $m_3$ & $m_4$ & $m_5$ & $m_6$ & $\Phi_1$ & $\Phi_2$ & $\tilde{\Phi}_2$ & $\Phi_3$ & $\Phi_4$ & $\Phi_5$ & $\Phi_6$ & $X$ \\
\midrule
$\surd$ & 0 & 0 & 0 & 1 & 1 & 0 & 0 & 0 & 1 & 0 & 0 & 0 & 0 & 0 \\
 & 0 & 0 & 0 & 1 & 1 & 0 & 0 & 1 & 0 & 0 & 0 & 0 & 0 & 0 \\
$\surd$ & 0 & 0 & 1 & 0 & 0 & 0 & 0 & 0 & 1 & 0 & 0 & 0 & 0 & 0 \\
 & 0 & 0 & 1 & 0 & 0 & 0 & 0 & 1 & 0 & 0 & 0 & 0 & 0 & 0 \\
$\surd$ & 0 & 1 & 0 & 0 & 0 & 0 & 0 & 0 & 0 & 0 & 0 & 1 & 1 & 1 \\
$\surd$ & 0 & 1 & 0 & 0 & 0 & 0 & 1 & 0 & 0 & 1 & 0 & 0 & 0 & 1 \\
$\surd$ & 1 & 0 & 0 & 0 & 0 & 1 & 0 & 0 & 1 & 0 & 0 & 0 & 0 & 0 \\
 & 1 & 0 & 0 & 0 & 0 & 1 & 0 & 1 & 0 & 0 & 0 & 0 & 0 & 0 \\ 
\bottomrule
\end{tabular}
\caption{Half-BPS dressed monopole operators in the 3d theory for $(A_{1},D_{4})$. Operators included in the superpotential are indicated with a check mark.}
\label{tab:d4mono}
\end{table}

There are $8$ dressed monopole operators in total, with $R=2$ and neutral under $U(1)_{f}$, listed in Table~\ref{tab:d4mono}.
The global symmetry before turning on the superpotential is $U(1)^{6}$, and we expect that there exists a superpotential deformation that preserves $U(1)_A \times U(1)_{f_1} \times U(1)_{f_2} \subset SO(4)_R \times SU(3)$. We find that the superpotential shown in Table \ref{tab:d4mono} satisfies this requirement. We conjecture that with this superpotential, the theory flows to $\CN=4$ SCFT in the IR with enhanced $SU(3)$ flavor symmetry. 

We identify the flavor and axial symmetries as (up to permutation of $U(1)_{1,3,4}$ and $\Phi_{2}\leftrightarrow \tilde{\Phi}_{2}$) 
\begin{align}
\begin{split}
	U(1)_{f_{1}}&=~U(1)_{T_{1}}-U(1)_{T_{6}},\\
	U(1)_{f_{2}}&=~-U(1)_{T_{1}}-U(1)_{T_{4}}+U(1)_{T_{5}}+U(1)_{T_{6}},\\
	U(1)_A=~&U(1)_{T_1}+U(1)_{T_3}+U(1)_{T_4}-U(1)_{\tilde{\Phi}_2}
\end{split}
\end{align} 
where $U(1)_A \subset SO(4)_R$ and $U(1)_{f_1} \times U(1)_{f_2} \subset SU(3)$ are the Cartans of global symmetry. 

\begin{figure}[tb]
	\centering
	\includegraphics[width=0.65\textwidth]{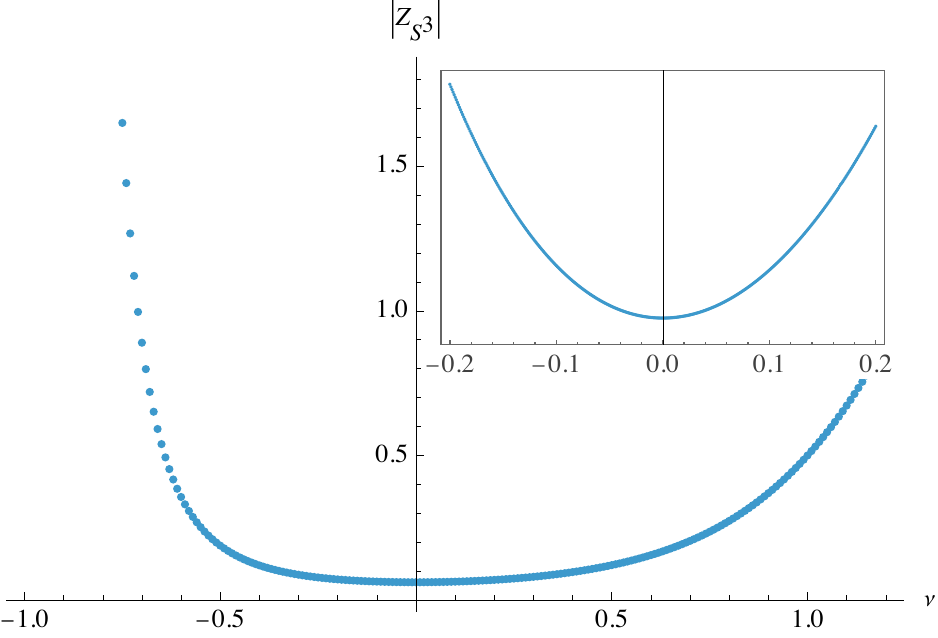}
	\caption{The absolute value of the three-sphere partition function for the 3d theory of $(A_{1},D_{4})$ as a function of the mixing parameter $\nu$ with $U(1)_{A}=U(1)_{\Phi_{2}}$. The function $|Z_{S^{3}}|$ is minimized at $\nu=0$.}
	\label{fig:d4Fmax}
\end{figure}

To obtain the superconformal $R$-symmetry of this theory, we perform F-maximization. We see that the three-sphere partition function is indeed minimized at $\nu=0$; see Figure~\ref{fig:d4Fmax}.

Now, the superconformal index and its Coulomb/Higgs limits for this theory can be computed to give
\begin{align}
\begin{split}
	\mathcal{I}_{SCI}(\mathfrak{q},T,z) &=1+T\chi^{SU(3)}_{[1,1]}\mathfrak{q}^{\frac{1}{2}}+\left(-1-\chi^{SU(3)}_{[1,1]}+T^2\chi^{SU(3)}_{[2,2]}\right)\mathfrak{q}\\
	&\qquad
    +\left( T(1-\chi^{SU(3)}_{[2,2]}-\chi^{SU(3)}_{[3,0]}-\chi^{SU(3)}_{[0,3]})+T^3\chi^{SU(3)}_{[3,3]}) \right)\mathfrak{q}^{\frac{3}{2}}+\cdots,\\
    \mathcal{I}_{C}(\mathfrak{q},z) &=1,\\
	\mathcal{I}_{H}(T,z) &=1+\chi_{[1,1]}^{SU(3)}T+\chi_{[2,2]}^{SU(3)}T^{2}+\chi_{[3,3]}^{SU(3)}T^{3}+\chi_{[4,4]}^{SU(3)}T^4+\cdots \ , 
\end{split}
\end{align}
where $\chi_R^{SU(3)}$ denotes the character for $SU(3)$ in representation $R$ and we omitted the fugacity $z=(z_1, z_2)$. 
The term $\chi^{SU(3)}_{[1,1]}\mathfrak{q}^{\frac{1}{2}}$ in the superconformal index comes from the $SU(3)_{f}$ flavor symmetry, and the extra $-\mathfrak{q}$ corresponds to the $\mathcal{N}=4$ stress-energy tensor.
The Coulomb and Higgs limits correctly reproduce the Hilbert series of the trivial Coulomb branch and the Higgs branch of  $\mathcal{M}_{SU(3)}$, which is given by the minimal nilpotent orbit of $SU(3)$. The Higgs branch of $(A_1, D_4)$ theory is given by the centered one $SU(3)$-instanton moduli space, or minimal nilpotent orbit of $SU(3)$. Its Hilbert series is given as \cite{Benvenuti:2010pq, Keller:2011ek}
\begin{align}
	\mathrm{HS}(\mathcal{M}_{SU(3)};T, z)=\sum _{i=0}^{\infty }\chi^{SU(3)}_{[i,i]}T^{i} \ . 
\end{align}

The $T$-refined half-index in the A-twisted limit matches our conjectured formula for the Macdonald index of $(A_{1},D_{4})$ AD theory:
\begin{align} \label{eq:d4Mac}
\begin{split}
	\II&=\frac{1}{(q)_\infty^{6}}\sum _{m\in\mathbb{Z} ^{6}}q^{\frac{1}{2}\sum _{i,j=1}^{4}b_{ij}m_{i}m_{j}}q^{\sum _{i=1}^{4}m_{i}}\left(\frac{z_{1}^2}{z_{2}}\right)^{m_{4}-m_{6}}\left(\frac{z_{1}}{z_{2}^2}\right)^{-m_{3}-m_{4}+m_{5}+m_{6}}T^{m_{1}+m_{3}+m_{4}}\\
	&\qquad \qquad \times \left(\prod _{i=1}^{6}(q^{1+m_{i}})_\infty\right)(q^{1+m_{2}}T)_\infty(q^{1+m_{1}+m_{3}+m_{4}-m_{5}-m_{6}})_\infty\\
	&=(q)_\infty(qT)_\infty \sum_{m\in\mathbb{Z}^{6}} \frac{q^{\frac{1}{2} m^t b m}q^{\sum _{i=1}^{4}m_{i}}\left(\frac{z_{1}^2}{z_{2}}\right)^{m_{4}-m_{6}}\left(\frac{z_{1}}{z_{2}^2}\right)^{-m_{3}-m_{4}+m_{5}+m_{6}}T^{m_{1}+m_{3}+m_{4}}}
	 {\left(\prod _{i=1}^{6}(q)_{m_{i}}\right)(qT)_{m_{2}}(q)_{m_{1}+m_{3}+m_{4}-m_{5}-m_{6}}}. 
\end{split}
\end{align}
Upon expanding in $q$-series, we obtain
\begin{align}
\begin{split}
    \II &=1+qT\chi^{SU(3)}_{[1,1]}+q^2\left(T(1+\chi^{SU(3)}_{[1,1]})+T^2\chi^{SU(3)}_{[2,2]}  \right)\\
	 &\qquad +q^3\left( T(1+\chi^{SU(3)}_{[1,1]})+T^2(\chi^{SU(3)}_{[1,1]}+\chi^{SU(3)}_{[3,0]}+\chi^{SU(3)}_{[0,3]}+\chi^{SU(3)}_{[2,2]})+T^3\chi^{SU(3)}_{[3,3]} \right)+\cdots , 
\end{split}
\end{align}
which agrees with the known expression. 

\subsubsection{$D_{6}$}
The formula \eqref{eq:d2nSchur} gives
\begin{align}
	I_{\text{Schur}}^{D_{6}}=
	(q)_\infty^{4}\sum _{m\in \mathbb{Z} ^{8}}
	\frac{
	q^{\frac{1}{2} m^t b^{D_{6}}m }(q^{\frac{1}{2}})^{ m_{1}+m_{7}+2\sum _{i=2}^{6}m_{i}}(w^2)^{m_{1}-m_{5}-m_{7}+m_{8}}(\frac{x}{w})^{m_{1}-m_{7}}
	}
	{
		\left(\prod _{i=1}^{8}(q)_{m_{i}}\right)\left(\prod _{j=2,4}(q)_{m_{j}}\right)(q)_{m_{1}+m_{3}-m_{7}}(q)_{-m_{1}+m_{5}+m_{6}+m_{7}-m_{8}}
	} \ , 
\end{align} 
from which we can read off the ACSM theory with $U(1)^8$ gauge group and 12 chiral multiplets. 

Before turning on the superpotential, this theory has $U(1)^{12}$ flavor symmetry. We look for an appropriate superpotential that breaks all the symmetry except for $U(1)_{A}\times  U(1)_{f_{1}}\times U(1)_{f_{2}}$. 
Since this theory has a rather high rank, we were not able to search over all possible monopole combinations for superpotential deformation. But we do find a good candidate superpotential given as 
\begin{align}
\begin{split}
	\mathcal{W} &=
V_{1,7}\Phi_{2}+V_{3}\Phi_{2}\Phi_{4}+V_{3}\tilde{\Phi}_{2}\tilde{\Phi}_{4}  +V_{6}\tilde{\Phi}_{4}+V_{5,8}\tilde{\Phi}_{4} \\
	&\quad +V_{2}\Phi_{1}\Phi_{3}+V_{2}\Phi_{7}\hat{\Phi}_{3}+V_{4}\Phi_{3}\Phi_{5}\Phi_{6}+V_{4}\Phi_{8}\hat{\Phi}_{3}X \ . 
\end{split}	
\end{align} 
We conjecture that upon turning on this superpotential, the ACSM theory flows to $\CN=4$ SCFT in the IR with $SU(2)\times U(1)$ flavor symmetry. 
We identify the axial symmetry $U(1)_A \subset SO(4)_R$ as
\begin{align} 
 U(1)_A=\frac12(U(1)_{T_1}+U(1)_{T_7}) + U(1)_{T_3}+U(1)_{T_5}+U(1)_{T_6} - U(1)_{\Phi_2}-U(1)_{\tilde{\Phi}_4}.
\end{align}
Since this theory has a rather high rank, we were not able to find a superconformal index to a reasonable order to test the SUSY enhancement, and the Coulomb/Higgs moduli space. 
However, we do find that the $U(1)_A$ refined half-index of our theory indeed reproduces the Macdonald index of $(A_1, D_6)$ theory: 
\begin{align} \label{eq:d6Mac}
\begin{split}
	\II(&q,T,w,x)\\
	&=\frac{1}{(q)_\infty^{8}}\sum _{m\in\mathbb{Z} ^{8}}q^{\frac{1}{2} m^t b^{D_{6}}m}(q^{\frac{1}{2}})^{m_{1}+m_{7}+2\sum _{i=2}^{6}m_{i}}
    (w^2)^{m_{1}-m_{5}-m_{7}+m_{8}}\left(\frac{x}{w}\right)^{m_{1}-m_{7}}T^{m_{3}+m_{5}+m_{6}+\frac{1}{2}(m_{1}+m_{7})}\\
	&\quad\times \left( \prod _{i=1}^{8}(q^{1+m_{i}})_\infty \right)(Tq^{1+m_{2}})_\infty(Tq^{1+m_{4}})_\infty(q^{1+m_{1}+m_{3}-m_{7}})_\infty(q^{1-m_{1}+m_{5}+m_{6}+m_{7}-m_{8}})_\infty\\
	&=(q)_\infty^2(qT)_\infty^2\\
    &\times\sum _{m\in \mathbb{Z} ^{8}}
	\frac{
	q^{\frac{1}{2} m^t b^{D_{6}} m}(q^{\frac{1}{2}})^{ m_{1}+m_{7}+2\sum _{i=2}^{6}m_{i}}(w^2)^{m_{1}-m_{5}-m_{7}+m_{8}}(\frac{x}{w})^{m_{1}-m_{7}}T^{m_{3}+m_{5}+m_{6}+\frac{1}{2}(m_{1}+m_{7})}
	}
	{
		\left(\prod _{i=1}^{8}(q)_{m_{i}}\right)\left(\prod _{j=2,4}(qT)_{m_{j}}\right)(q)_{m_{1}+m_{3}-m_{7}}(q)_{-m_{1}+m_{5}+m_{6}+m_{7}-m_{8}}
	}
\end{split}
\end{align}
Indeed, the half-index reproduces our conjectured fermionic sum formula. Upon series expansion, we obtain 
\begin{align}
\begin{split}
\II(q,T,w,x)
	&=1+q T \left(w^2+\frac{1}{w^2}+2\right)+q^{3/2} T^{3/2} \left(w x+\frac{w}{x}+\frac{1}{w x}+\frac{x}{w}\right)\\
	 &\qquad +q^2 \left(T^2 \left(w^4+\frac{1}{w^4}+2 w^2+\frac{2}{w^2}+3\right)+T \left(w^2+\frac{1}{w^2}+3\right)\right)\\
	 &\qquad +q^{5/2} \Bigg(T^{5/2}\left(w^3 x+\frac{w^3}{x}+\frac{x}{w^3}+\frac{1}{w^3 x}+2 w x+\frac{2 w}{x}+\frac{2}{w x}+\frac{2 x}{w} \right)\\
	 &\qquad \qquad \qquad +T^{3/2} \left(w x+\frac{w}{x}+\frac{1}{w x}+\frac{x}{w}\right) \Bigg) + \cdots \ , 
\end{split}
\end{align} 
which agrees with the known result for the Macdonald index. 
This is consistent with the expectation that the bulk theory has enhanced $\CN=4$ supersymmetry. 

\subsubsection{General Form}
Based on the expressions \eqref{eq:d4Mac} and \eqref{eq:d6Mac}, we consider the following superpotential
\begin{align}
\begin{split}
\mathcal{W}&=V_{1,2n+3}\Phi_{2}+V_{2}\Phi_{2n+3}\hat{\Phi}_{3}\\
	&\quad +\sum _{i=1}^{n-1}V_{2i}\Phi_{2i-1}\Phi_{2i+1}+\sum _{i=2}^{n-1}V_{2i}\hat{\Phi}_{2i-1}\hat{\Phi}_{2i+1}\\
	&\quad +\sum _{i=1}^{n-1}V_{2i+1}\Phi_{2i}\Phi_{2i+2}+\sum _{i=1}^{n-1}V_{2i+1}\tilde{\Phi}_{2i}\tilde{\Phi}_{2i+2}\\
	&\quad +V_{2n}\Phi_{2n-1}\Phi_{2n+1}\Phi_{2n+2}+V_{2n}\hat{\Phi}_{2n-1}\Phi_{2n+4}X+V_{2n+2}\tilde{\Phi}_{2n}+V_{2n+1,2n+4}\tilde{\Phi}_{2n} \ . 
\end{split}
\end{align}
Upon this superpotential deformation, we conjecture that the theory flows to a $\mathcal{N}=4$ SCFT in the IR along with a flavor symmetry enhancement to $SU(3)$ for $n=1$ and $SU(2)\times U(1)$ for $n>1$. We also identify the axial $U(1)_A \subset SO(4)_R$ as follows: 
\begin{align}
U(1)_A=
\begin{cases}
    \displaystyle \sum_{i=1}^{n+1} U(1)_{T_{2i-1}}+U(1)_{T_{2n+2}}-\sum_{i=1}^{n-2}\left(U(1)_{\tilde\Phi_{2i}}-U(1)_{\Phi_{2i+2}}\right)-U(1)_{\tilde\Phi_{2n}} & \text{($n$ odd)}\\
    \displaystyle \frac12\left(U(1)_{T_1}+U(1)_{T_{2n+3}}\right)+\sum_{i=2}^{n+1} U(1)_{T_{2i-1}}+U(1)_{T_{2n+2}}-\sum_{i=1}^{n-1}\left(U(1)_{\Phi_{2i}}-U(1)_{\tilde\Phi_{2i+2}}\right) & \text{($n$ even)}\\  
\end{cases}
\end{align}
With this at hand, we compute the $U(1)_A$-refined half-index given in a fermionic sum formula we propose
\begin{align}
\begin{split}
	I_{\text{Mac}}^{D_{2n+2}}&(q, T, z) = (q)_{\infty }^{n}(qT)_\infty^n \\
	& \times
	\sum _{k,\ell \in \mathbb{Z} ^{2n+2}} 
	\frac{
	q^{\frac{1}{2}k^{t}b^{D_{2n+2}}\ell}
	q^{\frac12\sum _{i}^{2n+2}(k_{i}+\ell_{i})}}
	{(q)_{k_{2n+1}}(q)_{k_{2n+2}}(q)_{\ell_{2n+1}}(q)_{\ell_{2n+2}}\prod_{i=1}^{n}(q)_{k_{2i-1}}(q)_{\ell_{2i-1}}\prod _{j=1}^{n}(qT)_{k_{2j}}(q)_{\ell_{2j}}}\\
    &\quad \times(\mathrm{Tr} [X_{\gamma_{f_{1}}}])^{\ell_{2n+2}-k_{2n+2}}(\mathrm{Tr} [X_{\gamma_{f_{2}}}])^{\ell_{1}-k_{1}}
	(T^{\frac{1}{2}})^{\sum _{i=1}^{n+1}(k_{2i-1}+\ell_{2i-1})+k_{2n+2}+\ell_{2n+2}}\\
	&\quad \times
	\left(\prod_{i=1}^{n}\delta_{k_{2i},\ell_{2i}}\right)
	\left(\prod _{j=1}^{n-1}\delta_{(-1)^{j+1}k_{1}+k_{2j+1},(-1)^{j+1}\ell_{1}+\ell_{2j+1}}\right)\\
	&\quad \times \delta_{(-1)^{n+1}k_{1}+k_{2n+1}+k_{2n+2},(-1)^{n+1}\times \ell_{1}+\ell_{2n+1}+\ell_{2n+2}}.
\end{split}
\end{align} 
This formula indeed agrees with the Macdonald index of the $(A_{1},D_{2n+2})$ AD theory, and the Macdonald grading in 4d corresponds to the $U(1)_{A}$ charge.

\subsection{$(A_1, E_n)$ theories}
\subsubsection{$E_6$}

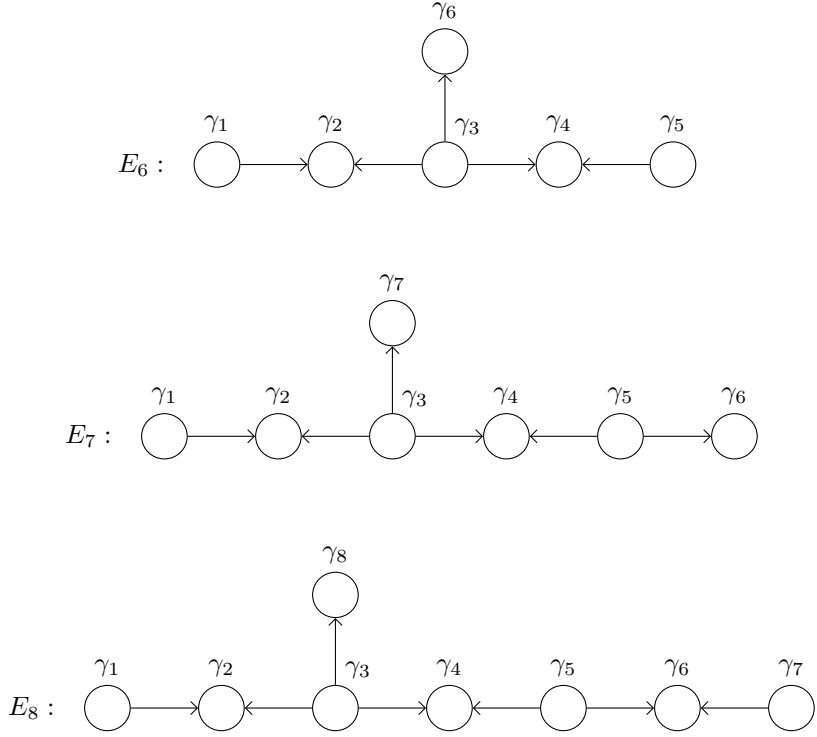
\begin{figure}
\centering
\begin{tikzpicture}
    \node[] (0) at (-1,0){$E_6:$};
    \node[W] (1) at (0,0){};
    \node[above=3mm] at (1) {$\gamma_1$};
    
    \node[W] (2) at (1.5,0){};
    \node[above=3mm] at (2) {$\gamma_2$};
    
    \node[W] (3) at (3,0){};
    \node[shift={(3mm,5mm)}] at (3) {$\gamma_3$};
    
    \node[W] (4) at (4.5,0){};
    \node[above=3mm] at (4) {$\gamma_4$};
    
    \node[W] (5) at (6,0){};
    \node[above=3mm] at (5) {$\gamma_5$};
    
    \node[W] (6) at (3,1.5){};
    \node[above=3mm] at (6) {$\gamma_6$};
    
    \draw[->] (1)--(2);
    \draw[<-] (2)--(3);
    \draw[->] (3)--(4);
    \draw[<-] (4)--(5);
    \draw[->] (3)--(6);
\end{tikzpicture}
\vspace{1cm}

\begin{tikzpicture}
    \node[] (0) at (-1,0){$E_7:$};
    \node[W] (1) at (0,0){};
    \node[above=3mm] at (1) {$\gamma_1$};

    \node[W] (2) at (1.5,0){};
    \node[above=3mm] at (2) {$\gamma_2$};
    
    \node[W] (3) at (3,0){};
    \node[shift={(3mm,5mm)}] at (3) {$\gamma_3$};
    
    \node[W] (4) at (4.5,0){};
    \node[above=3mm] at (4) {$\gamma_4$};
    
    \node[W] (5) at (6,0){};
    \node[above=3mm] at (5) {$\gamma_5$};
    
    \node[W] (6) at (7.5,0){};
    \node[above=3mm] at (6) {$\gamma_6$};
    
    \node[W] (7) at (3,1.5){};
    \node[above=3mm] at (7) {$\gamma_7$};
    
    \draw[->] (1)--(2);
    \draw[<-] (2)--(3);
    \draw[->] (3)--(4);
    \draw[<-] (4)--(5);
    \draw[->] (5)--(6);
    \draw[->] (3)--(7);
\end{tikzpicture}

\vspace{1cm}

\begin{tikzpicture}
    \node[] (0) at (-1,0){$E_8:$};
    \node[W] (1) at (0,0){};
    \node[above=3mm] at (1) {$\gamma_1$};
    
    \node[W] (2) at (1.5,0){};
    \node[above=3mm] at (2) {$\gamma_2$};
    
    \node[W] (3) at (3,0){};
    \node[shift={(3mm,5mm)}] at (3) {$\gamma_3$};
    
    \node[W] (4) at (4.5,0){};
    \node[above=3mm] at (4) {$\gamma_4$};
    
    \node[W] (5) at (6,0){};
    \node[above=3mm] at (5) {$\gamma_5$};
    
    \node[W] (6) at (7.5,0){};
    \node[above=3mm] at (6) {$\gamma_6$};
    
    \node[W] (7) at (9,0){};
    \node[above=3mm] at (7) {$\gamma_7$};
    
    \node[W] (8) at (3,1.5){};
    \node[above=3mm] at (8) {$\gamma_8$};
    
    \draw[->] (1)--(2);
    \draw[<-] (2)--(3);
    \draw[->] (3)--(4);
    \draw[<-] (4)--(5);
    \draw[->] (5)--(6);
    \draw[<-] (6)--(7);
    \draw[->] (3)--(8);
  \end{tikzpicture}
  \caption{\label{fig:e678quiver}
  BPS quiver for the $(A_1, E_6),(A_1, E_7)$ and $(A_1, E_8)$ theory in the canonical chamber}
\end{figure}

The IR trace formula for the Schur index of the $(A_1, E_6)$ theory gives
\begin{align}
I_{\text{Schur}}^{E_6}(q)=(q)_\infty^6\sum_{m\in \mathbb{Z}^6}
\frac{q^{\frac12 \sum_{i,j=1}^6 b^{E_6}_{ij}m_im_j}q^{\sum_{i=1}^6 m_i}}{\prod_{i=1}^n(q)_{m_i}^2}.
\end{align}
As before, from this we find the 3d $U(1)^6$ Chern-Simons theory with level $K= b^{E_6}$  and the mixed Chern-Simons levels between the gauge $U(1)$s and $R$-symmetry $\rho=\{2,2,\cdots,2\}$.
The theory is coupled to $12$ chiral multiplets $\Phi_i$ and $\tilde\Phi_i$ for $i=1,2,\cdots,6$, all with $R$-charge $0$, and their gauge charges are $-1$ under $U(1)_i$.

From our conjectural form of the Macdonald index in section \ref{sec: conjectural Macdonald}, we conjecture that the $U(1)_A$ symmetry can be identified with
\begin{align}
    U(1)_A^{E_6}&=U(1)_{T_2}+U(1)_{T_4}+U(1)_{T_6}-U(1)_{\Phi_1}-U(1)_{\tilde{\Phi}_3}-U(1)_{\Phi_5},
\end{align}
and the superpotential deformation as
\begin{align}
\begin{split}
    \mathcal{W}^{E_6}&=
 V_1\widetilde{\Phi}_2+V_2\Phi_1\Phi_3+V_3\Phi_2\widetilde{\Phi}_4\widetilde{\Phi}_6+V_3\widetilde{\Phi}_2\widetilde{\Phi}_4\Phi_6+V_3\widetilde{\Phi}_2\widetilde{\Phi}_4\widetilde{\Phi}_6\\
    &\qquad +V_4\Phi_3\Phi_5+V_4\widetilde{\Phi}_3\widetilde{\Phi}_5+V_5\Phi_4+V_5\widetilde{\Phi}_4+V_6\widetilde{\Phi}_3\ .
\end{split}
\end{align}
Indeed, upon computing the $T$-refined half-index with A-twist, we recover
\begin{align}
\begin{split}
    \II^{E_6}(q,T)&=\frac{1}{(q)_\infty^6}\sum_{m\in \mathbb{Z}^6}q^{\frac{1}{2}\sum_{i,j=1}^6b_{ij}^{E_6}m_im_j}q^{\sum_{i=1}^6 m_i}T^{m_2+m_4+m_6}\\
    &\qquad \times\left(\prod_{i=1,3,5}(q^{1+m_i}T)_\infty(q^{1+m_i})_\infty\right) \left(\prod_{j=2,4,6}(q^{1+m_i})_\infty^2\right)\\
    &=(q)_\infty^3(qT)_\infty^3\sum_{m\in \mathbb{Z}^6}
    \frac{q^{\frac{1}{2}\sum_{i,j=1}^6b_{ij}^{E_6}}q^{\sum_{i=1}^6 m_i}T^{m_2+m_4+m_6}}{\left(\prod_{i=1,3,5}(qT)_{m_i}(q)_{m_i}\right)\left(\prod_{i=2,4,6}(q)_{m_i}^2\right)} \ , 
\end{split}
\end{align}
which can be expanded in $q$-series to obtain
\begin{align}
\II^{E_6} =1 + T q^2 + (T + T^2) q^3 + (T + 2 T^2) q^4 + (T + 2 T^2) q^5 + (T + 3 T^2 + 2 T^3) q^6+\cdots . 
\end{align}

\paragraph{Alternative realization}
Alternatively, we can use another fermionic sum representation of the character of the $W_{3}$ minimal model $W_{3}(3,7)$, which is known to be a chiral algebra associated to the $E_6$ theory. From \cite{Feigin:2005gz}, we have
\begin{align}
\chi^{3,7}(q) = \sum_{n_{1,2, 3, 4} \ge 0} \frac{q^{(n_1+n_2+n_3)^2 + (n_2+n_3)^2 + n_3^2 + n_4^2 + (n_1 + 2n_2 + 3n_3)n_4 + (n_1 + 2n_2 + 3n_3 +2n_4)}}{(q)_{n_1} (q)_{n_2}(q)_{n_3}(q)_{n_4}} \ ,
\end{align}
which can be understood as the half-index of a 3d $U(1)^4$ Chern-Simons theory with level
\begin{align}
    K=
	\begin{pmatrix}
		2&2&2&1\\2&4&4&2\\2&4&6&3\\1&2&3&1
	\end{pmatrix} \ . 
\end{align}
The mixed Chern-Simons levels between the gauge $U(1)$s and $R$-symmetry are $\rho=\{2,4,6,4\}$. There are 4 chiral multiplets  $\Phi_i$ for $i=1,2,3,4$, all with trial $R$-charge 0 and gauge charge $-1$ under $U(1)_i$.

\begin{figure}[bt]
    \centering
	\includegraphics[width=0.65\textwidth]{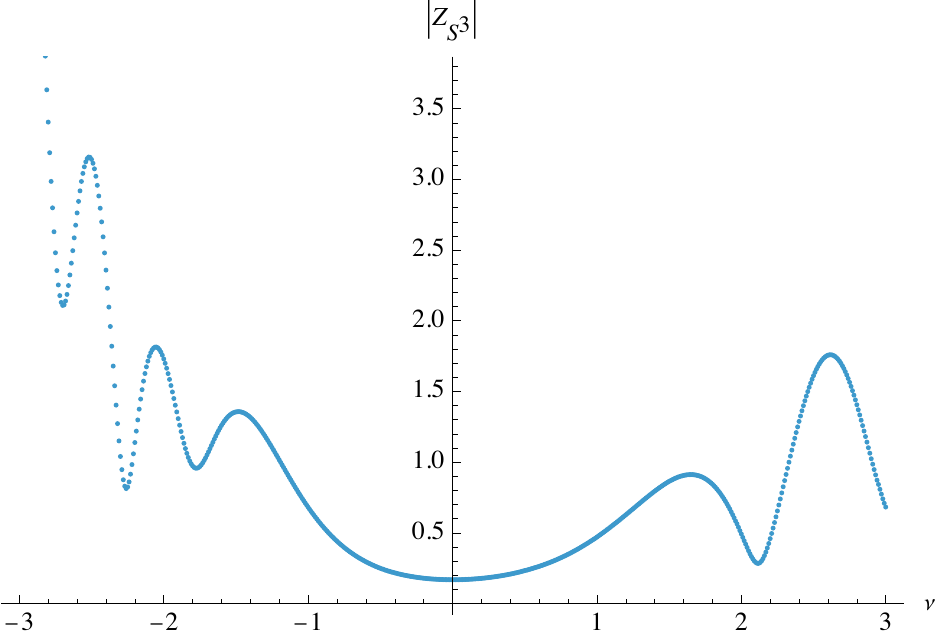}
	\caption{The absolute value of three-sphere partition function for the 3d theory from \eqref{eq:refinedw37} as a function of the mixing parameter $\nu$ with $U(1)_{A}=U(1)_{T_1}+2U(1)_{T_2}+3U(1)_{T_3}+2U(1)_{T_4}$. The function $|Z_{S^{3}}|$ is minimized at $\nu=0$}
	\label{fig:e6Fmax}
\end{figure}

One can verify that there exist four half-BPS gauge-invariant monopole operators:
\begin{align}
	V_{\{1,-1,1,-1\}},V_{\{-1,2,-1,0\}},V_{\{2,-1,0,0\}},V_{\{0,1,-2,2\}}\ ,
\end{align} 
where $V_{\{m_1,m_2,m_3,m_4\}}$ denotes bare monopole with magnetic charge $m_i$ corresponding to $U(1)_i$.
If we choose to deform with 
\begin{align}
    \mathcal{W}=V_{\{1,-1,1,-1\}}+V_{\{2,-1,0,0\}}+V_{\{0,1,-2,2\}}, 
\end{align}
there remains $U(1)^2$ unbroken global symmetry $U(1)_{T_{1}}+2U(1)_{T_{2}}-U(1)_{T_{4}}$ and $U(1)_{T_{3}}+U(1)_{T_{4}}$.
For any other choice of three monopole deformations, only one $U(1)$ global symmetry remains unbroken, and the surviving global symmetry always takes the form of
\begin{align}\label{eq:w37Axial}
U(1)_{A}=U(1)_{T_{1}}+2U(1)_{T_{2}}+3U(1)_{T_{3}}+2U(1)_{T_{4}}\ .
\end{align}
Performing $F$-maximization with this $U(1)_A$ (see Figure~\ref{fig:e6Fmax}), and calculating the superconformal index, 
We find
\begin{align}
\begin{split}
	\mathcal{I}_{SCI}(\mathfrak{q},T)&=1-\mathfrak{q}+T^{-1}\mathfrak{q}^{3/2}+(-1+T^2)\mathfrak{q}^2-2T\mathfrak{q}^{\frac{5}{2}} \\
    &\qquad+(3+T^{-2}+T^2) \mathfrak{q}^3-(3 T^{-1}+5T)\mathfrak{q}^{\frac{7}{2}}+\cdots ,\\
    \mathcal{I}_{C}(\mathfrak{q})&=1\ ,\\
	\mathcal{I}_{H}(T)&=1\ ,
\end{split}
\end{align}
which provides a strong consistency check of our proposal. Computing the refined A-twisted half-index with this symmetry, we find
\begin{align}\label{eq:refinedw37}
	\chi^{3,7}(q,T) = \sum_{n_{1,2, 3, 4} \ge 0} \frac{q^{(n_1+n_2+n_3)^2 + (n_2+n_3)^2 + n_3^2 + n_4^2 + (n_1 + 2n_2 + 3n_3)n_4 + (n_1 + 2n_2 + 3n_3 +2n_4)}}{(q)_{n_1} (q)_{n_2}(q)_{n_3}(q)_{n_4}} T^{n_1 + 2n_2 +3n_3 + 2n_4}\ ,
\end{align}
which indeed agrees with the Macdonald index computed in \cite{Foda:2019guo}\ .

\subsubsection{$E_7$}

For $(A_1, E_7)$ theory, the IR trace formula for the Schur index reads
\begin{align}
    I_\text{Schur}^{E_7}(q,z)=(q)_\infty^6\sum_{m\in\mathbb{Z}^8}\frac{q^{\frac12\sum_{i,j=1}^7b_{ij}^{E_7}m_im_j}(q^{\frac12})^{\sum_{i=1}^62m_i+m_7+m_8}z^{-m_7+m_8}}{\left(\prod_{i=1}^8(q)_{m_i}\right)\left(\prod_{j=1,2,3,5}(q)_{m_j}\right)(q)_{m_4+m_7-m_8}(q)_{m_6-m_7+m_8}}\ .
\end{align}
 From this we extract a 3d $U(1)^{8}$ Chern-Simons theory with level
\begin{align}
    K=\left( 
	\begin{array}{ccc|c}
		\multicolumn{3}{c}{\multirow{3}{*}{\raisebox{-1mm}{\scalebox{1.1}{$b^{E_7}$}}}}\vrule &0\\
		& & &\vdots\\
		& & & 0\\
		\hline
		0& \cdots  & 0 & 0
	\end{array}
	\right) \ , 
\end{align}
and the mixed Chern-Simons levels between the gauge $U(1)$s and the $R$-symmetry are $\rho=\{2,2,2,2,2,2,1,1\}$.
The theory is coupled to $14$ chiral multiplets, all with trial $R$-charge $0$.
Their gauge charges are
\begin{itemize}
\item $\Phi_i$ has charge $-1$ under $U(1)_i$ for $i=1,2,3,\cdots,8$.
\item $\tilde{\Phi}_j$ has charge $-1$ under $U(1)_j$ for $j=1,2,3,5$.
\item $X$ has charge $-1$ under $U(1)_4$, $-1$ under $U(1)_7$ and $+1$ under $U(1)_8$.
\item $Y$ has charge $-1$ under $U(1)_6$, $+1$ under $U(1)_7$ and $-1$ under $U(1)_8$.
\end{itemize}
The theory has $U(1)$ flavor symmetry, which corresponds to
\begin{align}
\begin{split}
    U(1)_f=-U(1)_{T_7}+U(1)_{T_8}.
\end{split}
\end{align}
Based on the conjectural formula for the Macdonald index in section \ref{sec: conjectural Macdonald}, we conjecture that the theory is subject to the following superpotential deformation
\begin{align}
    \mathcal{W}^{E_7}&=V_1\widetilde{\Phi}_2+V_2\Phi_1\Phi_3+V_2\widetilde{\Phi}_1\widetilde{\Phi}_3+V_3\Phi_2\Phi_8X+V_3\widetilde{\Phi}_2\Phi_4\Phi_7+V_3\widetilde{\Phi}_2\Phi_8X\notag\\
    &\qquad +V_4\Phi_3\Phi_5+V_4\widetilde{\Phi}_3\widetilde{\Phi}_5+V_5\Phi_4\Phi_6+V_5 XY+V_6\widetilde{\Phi}_5+V_{7,8}\Phi_3.
\end{align}
which leaves the $U(1)_f$ and the following unbroken symmetry, which we identify with the $U(1)_A$ symmetry:
\begin{align}
    U(1)_A=2(U(1)_{T_2}+U(1)_{T_4}+U(1)_{T_6})+U(1)_{T_7}+U(1)_{T_8}-2(U(1)_{\tilde{\Phi}_1}+U(1)_{\Phi_3}+U(1)_{\tilde{\Phi}_5}).
\end{align}
The refined A-twisted half-index reads 
\begin{align}
\begin{split}
    \II^{E_7}&=
    \frac{1}{(q)_\infty^8}\sum_{m\in\mathbb{Z}^8}q^{\frac{1}{2}\sum_{i,j=1}^7b^{E_7}_{ij}m_im_j}(q^{\frac{1}{2}})^{\sum_{i=1}^6 2m_i+m_7+m_8}(T^{\frac{1}{2}})^{2(m_2+m_4+m_6)+m_7+m_8}z^{-m_7+m_8}\\
    &~\times\left(\prod_{i=1}^8(q^{1+m_i})_\infty\right)\left(\prod_{j=1,3,5}(q^{1+m_j}T)_\infty\right)(q^{1+m_2})_\infty(q^{1+m_4+m_7-m_8})_\infty(q^{1+m_6-m_7+m_8})_\infty\\
    &=(q)_\infty^3(qT)_\infty^3 \\
    &\quad \cdot \sum_{m\in\mathbb{Z}^8}\frac{q^{\frac{1}{2}\sum_{i,j=1}^7b^{E_7}_{ij}m_im_j}(q^{\frac{1}{2}})^{\sum_{i=1}^62m_i+m_7+m_8}(T^{\frac{1}{2}})^{2(m_2+m_4+m_6)+m_7+m_8}z^{-m_7+m_8}}{\left(\prod_{i=1}^8(q)_{m_i}\right)\left(\prod_{j=1,3,5}(qT)_{m_j}\right)(q)_{m_2}(q)_{m_4+m_7-m_8}(q)_{m_6-m_7+m_8}} , 
\end{split}
\end{align}
which is our proposed form of the Macdonald index for $(A_1, E_7)$ theory in section \ref{sec: conjectural Macdonald}. Upon series expansion, we obtain
\begin{align}
\begin{split}
  \II^{E_7} &=
1 
+ qT 
+ q^{3/2}T^{3/2}\left(z + \frac{1}{z}\right) 
+ q^2(2T + T^2) 
+ q^{5/2}\left(T^{3/2}\left(z + \frac{1}{z}\right)+T^{5/2}\left(z + \frac{1}{z}\right)\right)\\
&\quad+ q^3\left(2T + 3T^2 + T^3\left(z^2 + \frac{1}{z^2} + 1\right)\right) \\
&\quad+ q^{\frac{7}{2}}\left(T^{3/2}\left(z + \frac{1}{z}\right) + T^{5/2}\left(3z + \frac{3}{z}\right) + T^{7/2}\left(z + \frac{1}{z}\right)\right) \\
&\quad+ q^4\left(2T + 6T^2 + T^3\left(z^2 + \frac{1}{z^2} + 3\right) + T^4\left(z^2 + \frac{1}{z^2} + 1\right)\right) \\
&\quad+ q^{\frac{9}{2}}\left(T^{\frac{3}{2}}\left(z + \frac{1}{z}\right) + T^{\frac{5}{2}}\left(5z + \frac{5}{z}\right) + T^{\frac{7}{2}}\left(3z + \frac{3}{z}\right) + T^{\frac{9}{2}}\left(z^3 + \frac{1}{z^3} + z + \frac{1}{z}\right)\right)\\
&\quad+ q^5\left(2T + 7T^2 + T^3\left(2z^2 + \frac{2}{z^2} + 8\right) + T^4\left(3z^2 + \frac{3}{z^2} + 3\right) + T^5\left(z^2 + \frac{1}{z^2} + 1\right)\right) \\
&\quad+\cdots .
\end{split}
\end{align}
As far as the authors are aware, there is no known explicit expression for the Macdonald index of the $(A_1, E_7)$ theory. This therefore provides a prediction for this theory.

\subsubsection{$E_8$}
The trace formula gives
\begin{align}
I_{\text{Schur}}^{E_{8}}(q)=(q)_\infty^8\sum_{m\in \mathbb{Z}^8}
\frac{q^{\frac12 \sum_{i,j=1}^8 b^{E_8}_{ij}m_im_j}q^{\sum_{i=1}^8 m_i}}{\prod_{i=1}^8(q)_{m_i}^2}.
\end{align}
As before, from this we find the 3d $U(1)^8$ Chern-Simons theory with level $K= b^{E_8}$  and the mixed Chern-Simons levels between the gauge $U(1)$s and $R$-symmetry $\rho=\{2,2,\cdots,2\}$.
There are $16$ chiral multiplets $\Phi_i$ and $\tilde\Phi_i$ for $i=1,2,\cdots,8$, all with $R$-charge $0$, and their gauge charges are $-1$ under $U(1)_i$

From our conjectural form of the Macdonald index in section \ref{sec: conjectural Macdonald}, we conjecture that the $U(1)_A$ symmetry is identified with
\begin{align}
    U(1)_A&=U(1)_{T_2}+U(1)_{T_4}+U(1)_{T_6}+U(1)_{T_8}-U(1)_{\Phi_1}-U(1)_{\tilde{\Phi}_3}-U(1)_{\Phi_5}-U(1)_{\tilde{\Phi}_7}, 
\end{align}
and the superpotential deformations as
\begin{align}
\begin{split}
   \mathcal{W}&=
 V_1\widetilde{\Phi}_2+V_2\Phi_1\Phi_3+V_3\Phi_2\widetilde{\Phi}_4\widetilde{\Phi}_8+V_3\widetilde{\Phi}_2\widetilde{\Phi}_4\Phi_8+V_3\widetilde{\Phi}_2\widetilde{\Phi}_4\widetilde{\Phi}_8\\
 &\qquad +V_4\Phi_3\Phi_5+V_4\widetilde{\Phi}_3\widetilde{\Phi}_5+V_5\Phi_4\widetilde{\Phi}_6+V_5\widetilde{\Phi}_4\widetilde{\Phi}_6+V_6\Phi_5\Phi_7+V_6\widetilde{\Phi}_5\widetilde{\Phi}_7\\
    &\qquad +V_7\Phi_6+V_7\widetilde{\Phi}_6+V_8\widetilde{\Phi}_3.
\end{split}
\end{align}

Computing the $T$-refined half-index with A-twist, we recover the conjecture form of the Macdonald index 
\begin{align}
\begin{split}
    \II^{E_8}(q,T)&=\frac{1}{(q)_\infty^8}\sum_{m\in \mathbb{Z}^8}q^{\frac{1}{2}\sum_{i,j=1}^8b_{ij}^{E_8}m_im_j}q^{\sum_{i=1}^8 m_i}T^{m_2+m_4+m_6+m_8}\\
    &\qquad \times \left(\prod_{i=1,3,5,7}(q^{1+m_i}T)_\infty(q^{1+m_i})_\infty\right) \left(\prod_{j=2,4,6,8}(q^{1+m_i})_\infty^2\right)\\
    &=(q)_\infty^4(qT)_\infty^4\sum_{m\in \mathbb{Z}^8}
    \frac{q^{\frac{1}{2}\sum_{i,j=1}^8b_{ij}^{E_8}m_im_j}q^{\sum_{i=1}^8 m_i}T^{m_2+m_4+m_6+m_8}}{\left(\prod_{i=1,3,5,7}(qT)_{m_i}(q)_{m_i}\right)\left(\prod_{i=2,4,6,8}(q)_{m_i}^2\right)} \ . 
\end{split}
\end{align}
Upon expanding in $q$, we obtain
\begin{align}
\begin{split}
    &=1+Tq^2+(T+T^2)q^3+(T+2T^2)q^{4}+(T+2T^2+T^3)q^{5}+(T+3T^2+3T^3)q^{6}+\cdots , 
\end{split}
\end{align}
which agrees with the known expressions for the Macdonald index. 

\paragraph{Alternative realization}
Alternatively, we can use another fermionic sum representation of the character of the $W_3$ minimal model $W_3(3,8)$, which is known to be a chiral algebra associated with the $E_8$ theory.
From \cite{andrews1999a2}, we have
\begin{align}\label{eq:w38t34}
\begin{split}
    \chi^{3,8}(q)
    &=\frac{(q,q,q^2,q^6,q^7,q^7,q^8,q^8;q^8)_\infty}{(q;q)_\infty^2}\\
    &=(q;q)_\infty \sum_{l_1,l_2,l_3,l_4 \in\mathbb{Z} }\frac{q^{l_1^2-l_1l_2+l_2^2+l_3^2+l_3l_4+l_4^2+l_1+l_2+l_3+l_4}}{(q)_{l_1-l_3}(q)_{l_2-l_4}(q)_{l_3}(q)_{l_4}(q)_{l_3+l_4+1}}\\
    &=(q;q)_\infty \sum_{l_1,l_2,k\in\mathbb{Z}}\frac{q^{l_1^2-l_1l_2+l_2^2+l_1+l_2+k^2+k}}{(q)_{l_1+l_2+1}(q)_{l_1-k}(q)_{l_2-k}(q)_{k}}
\end{split}
\end{align}

The last expression can be interpreted as an A-twisted half-index of the 3d $U(1)^3$ Chern-Simons theory with level
\begin{align}
    K=
	\begin{pmatrix}
		2&-1&0\\-1&2&0\\0&0&2
	\end{pmatrix}.
\end{align}
The mixed Chern-Simons levels between the gauge $U(1)$s and the $R$-symmetry are $\rho=\{0,0,-2\}$.
There are four chiral multiplets:
\begin{itemize}
    \item $\Phi_1$ has gauge charge $-1$ under $U(1)_1$ and $-1$ under $U(1)_2$. Its $R$-charge is $2$.
    \item $\Phi_2$ has gauge charge $-1$ under $U(1)_1$ and $+1$ under $U(1)_3$. Its $R$-charge is $0$.
    \item $\Phi_3$ has gauge charge $-1$ under $U(1)_2$ and $+1$ under $U(1)_3$. Its $R$-charge is $0$.
    \item $\Phi_4$ has gauge charge $-1$ under $U(1)_3$. Its $R$-charge is $2$.
\end{itemize}

The theory has precisely three half-BPS gauge-invariant monopole operators. Deforming by
\begin{align}
	\mathcal{W}=V_{\{-1,0,0,\}}\Phi_{3}^2\Phi_{4}+ V_{\{0,-1,0\}}\Phi_{2}^2\Phi_{4}+ V_{\{0,0,1\}}\Phi_{1}\ ,
\end{align}
there remains a single $U(1)$ global symmetry
\begin{align}
	U(1)_A=U(1)_{T_{1}}+U(1)_{T_{2}}+U(1)_{\Phi_{4}}.
\end{align}

\begin{figure}[bt]
	\centering
	\includegraphics[width=0.65\textwidth]{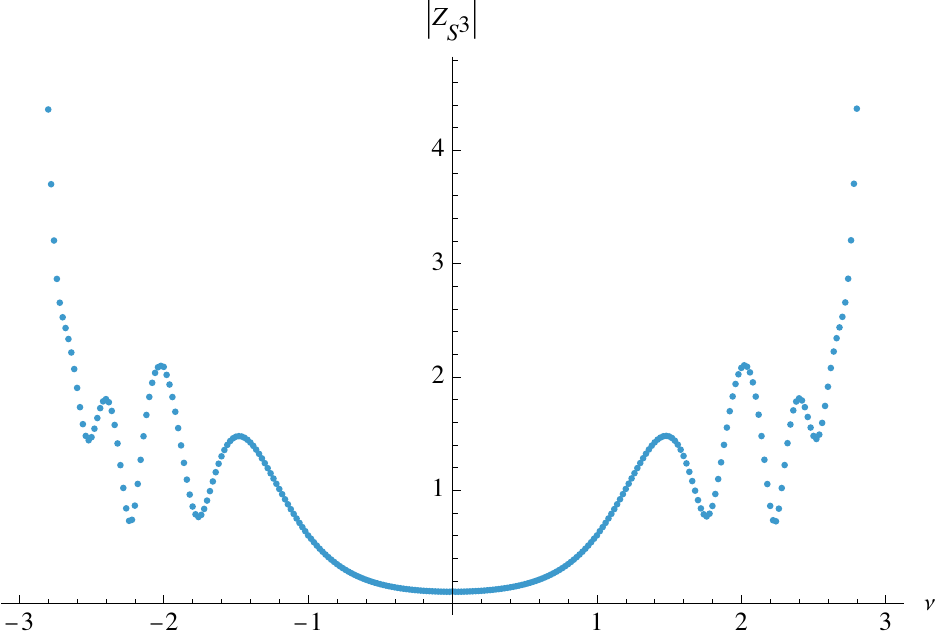}
	\caption{The absolute value of three-sphere partition function for the 3d theory from \eqref{eq:w38t34} as a function of the mixing parameter $\nu$ with $U(1)_{A}=U(1)_{\Phi_{2}}$. The function $|Z_{S^{3}}|$ is minimized at $\nu=0$}
	\label{fig:e8Fmax}
\end{figure}
The $F$-maximization with this $U(1)_A$ (see Figure~\ref{fig:e8Fmax}), and the superconformal index, which reads,
\begin{align}
\begin{split}
	\mathcal{I}_{SCI}(\mathfrak{q},T)&=1-\mathfrak{q}+(1+T^2+T^{-2})\mathfrak{q}^2+(3T^{-1}+3T)\mathfrak{q}^{\frac{5}{2}}+\cdots, \\
	\mathcal{I}_{C}(\mathfrak{q})&=1,\\
	\mathcal{I}_{H}(T)&=1 . 
\end{split}
\end{align}
provide a strong consistency check of our proposal.

For this example, in order to recover the Macdonald refinement, it requires additional mixing with the $U(1)^3_\partial$ symmetries at the boundary. Namely, we have to turn on the $U(1)_A$ fugacity with the specialization $x_i\rightarrow T$ for $i=1,2,3$,
 to recover the Macdonald index of the $(A_{1},E_{8})$ AD theory \cite{Song:2015wta,Watanabe:2019ssf} given as
\begin{align}
\begin{split}
	\II(q,T)&=\frac{1}{(q)_\infty^3}\sum _{m\in\mathbb{Z} ^{3}}q^{m_{1}^2-m_{1}m_{2}+m_{2}^2+m_{3}^2}q^{-m_{3}}T^{m_{1}+m_{2}+m_{3}}\\
	&\qquad\qquad\qquad\times(q^{m_{1}+m_{2}}T)_\infty(q^{1+m_{1}-m_{3}})_\infty(q^{1+m_{2}-m_{3}})_\infty(q^{m_{3}})_\infty\\
	&=qT^{3}\times \Big(1+Tq^2+(T+T^2)q^3+(T+2T^2)q^{4}+(T+2T^2+T^3)q^{5}\\
    &\qquad\qquad\quad +(T+3T^2+3T^3)q^{6}+(T+3T^2+4T^3)q^7+\cdots \Big) , 
\end{split}
\end{align} 
up to an overall factor.

\begin{acknowledgments}

This work is supported by the National Research Foundation of Korea (NRF) Grant RS-2024-00405629, as well as the KAIST-KIAS collaboration program. 
The work of Heeyeon Kim and Hongseok Kim is also supported by the National Research Foundation of Korea (NRF) Grant NRF2023R1A2C1004965 and by the POSCO Science Fellowship of POSCO TJ Park Foundation.
The work of J.S. is also supported by the National Research Foundation of Korea (NRF) Grant RS-2023-00208602, and also in part by the Walter Burke Institute for Theoretical Physics at Caltech and by the U.S. Department of Energy, Office of Science, Office of High Energy Physics, under Award Number DE-SC0011632.
\end{acknowledgments}


\appendix

\bibliographystyle{nb}
\bibliography{refs}

\end{document}